\documentclass[structabstract]{aa} 
\usepackage{natbib}
\bibpunct{(}{)}{;}{a}{,}{,}

\usepackage{graphicx}
\usepackage{longtable}
\usepackage{txfonts}
\begin{document}
   \title{Distances to dense cores that contain Very Low Luminosity Objects}
   \titlerunning{Distances to dense cores with VeLLOs}

   \author{
          Maheswar, G.\inst{1,2}
          ,
          Chang Won Lee\inst{2}
          ,
          Sami Dib\inst{3}
          }

   \institute{
   Aryabhatta Research Institute of Observational Sciences, Manora Peak, NainitaL263 129, India\\
   \email{maheswar@aries.res.in}
   \and
   Korea Astronomy \& Space Science Institute, 61-1 Hwaam-dong, Yusung-gu, Daejeon 305-348, Korea\\
   \email{cwl@kasi.re.kr}
   \and
   Astrophysics Group, Blackett Laboratory, Imperial College London, SW7 2AZ, United Kingdom.\\
   \email{s.dib@imperial.ac.uk}
             }

   \date{Received --- / Accepted ---}

   \abstract
   {}
   {To estimate distances to dense molecular cores that harbour Very Low Luminosity Objects (VeLLO) detected by Spitzer Space Telescope and to confirm their VeLLO nature.}
   {The cloud distances are estimated using near-IR photometric method. We use a technique that provides spectral classification of stars lying towards the fields containing the clouds into main sequence and giants. In this technique, the observed ($J-H$) and ($H-K_{s}$) colours are dereddened simultaneously using trial values of $A_{V}$  and a normal interstellar extinction law.  The best fit of the dereddened colours to the intrinsic  colours giving a minimum value of $\chi^{2}$ then yields the corresponding spectral type and $A_{V}$ for the star. The main sequence stars, thus classified, are then utilized in an $A_{V}$ versus distance plot to bracket the cloud distances. The typical error in the estimation of distances to the clouds are found to be $\sim18\%$.}
   {We estimated distances to seven cloud cores, IRAM04191, L1521F, BHR111, L328, L673-7, L1014, and L1148  using the above method. These clouds contain VeLLO candidates. The estimated distances to the cores are found to be $127\pm25$ pc (IRAM04191), $136\pm36$ pc (L1521F), $355\pm65$ pc (BHR111), $ 217\pm30 $ pc (L328), $ 240\pm45 $ pc (L673-7), $258\pm50 $ pc (L1014), and $301\pm55$ pc (L1148). We re-evaluated the internal luminosities of the VeLLOs discovered in these seven clouds using the distances estimated from this work. Except for L1014$-$IRS ($L_{int}=0.15$ $L_{\odot}$), all other VeLLO candidates are found to be consistent with the definition of a VeLLO ($L_{int}\leq0.1L_{\odot}$). In addition to the cores that harbour VeLLO candidates, we also obtained distances to the clouds L323, L675, L676, CB 188, L1122, L1152, L1155, L1157 and L1158 which are located in the directions of the above seven cores. Towards L1521F and L1148 we found evidence of the presence of multiple dust layers.}
   {}

   \titlerunning{Distances to dense cores with VeLLOs}
   \authorrunning{Maheswar et al.}
   \maketitle


\section{Introduction}

Some of the important physical parameters that are required to elucidate the still mysterious processes involved in the formation and evolution of both molecular clouds and pre-main sequence (PMS) stars depend crucially on the accurate determination of their distances.  But distances to most of the star forming and starless molecular clouds, especially those that are relatively isolated, are highly uncertain \citep{1995A&AS..113..325H}. Furthermore, the need to determine accurate distances to cloud cores can be understood from the recent discoveries of embedded sources having luminosities less than $0.1L_{\odot}$ called, very low luminous objects (VeLLOs), in a number of dense cores that were previously considered as starless \citep{2004ApJS..154..396Y, 2005AN....326..878K, 2006ApJ...651..945D, 2006ApJ...649L..37B, 2009ApJ...693.1290L, 2010ApJ...721..995D}. But, because the luminosity depends on distance squared, small errors in distance are effectively doubled. Therefore the above conclusions on the status of these objects as VeLLOs depend crucially on the distances to the parent clouds, which are highly uncertain (in some cases by over a factor of 2). For example, the authors who carried out studies on the VeLLO in L1014 have assumed a distance of 200 pc to the cloud . But recently, \citet{2006PASJ...58L..41M} have quoted a distance of 400-900 pc for L1014. Then the VeLLO, with luminosity of $L\sim0.09L_{\odot}$ (calculated using the assumed distance of 200 pc to the cloud), becomes $0.36-1.8 L_{\odot}$ and ceases to be a VeLLO.

Dense cores were previously considered to be starless if they do not contain an \textit{Infrared Astronomical Satellite (IRAS)} point source to a sensitivity of  $L\sim0.1L_{\odot}(d/140)^{2}$. But Spitzer observations of L1014, considered as a starless core, as a part of Spitzer Legacy program ``From Molecular Cores to Planet Forming Disks'' \citep[or c2d;][]{2003PASP..115..965E} came as a surprise! L1014 was found to contain an embedded source, L1014$-$IRS, with a very low luminosity of $L\sim0.09 L_{\odot}$ \citep{2004ApJS..154..396Y, 2005ApJ...633L.129B, 2006ApJ...640..391H}. Currently, the VeLLOs in fourteen more cores are identified in the full c2d sample \citep{2008ApJS..179..249D}. Of these, five of them (including L1014$-$IRS) have been studied in detail so far: IRAM 04191$+$1522 \citep{2006ApJ...651..945D}, L1521F$-$IRS \citep{2006ApJ...649L..37B, 2009ApJ...696.1918T}, L328$-$IRS \citep{2009ApJ...693.1290L}, L673-7$-$IRS \citep{2010ApJ...721..995D} and L1014$-$IRS \citep{2004ApJS..154..396Y, 2006ApJ...649L..37B, 2006ApJ...640..391H}. Sources with such low luminosities are hard to explain on the basis of the standard model of \citet{1987ARA&A..25...23S} as the observed luminosities are found to be more than an order of magnitude lower than what is expected from the model assuming a spherical mass accretion onto a protostellar object located on the stellar/substellar boundary \citep{2006ApJ...651..945D}. Thus, the discoveries of VeLLOs in dense cores can pose a formidable challenge to our  understanding of star formation. Recently, \citet{2010ApJ...710..470D} showed that the luminosity problem can be resolved and bring the model predictions in better agreement with the observations if the episodic mass accretion based on the simulation by \citet{2006ApJ...650..956V} is included in the evolutionary models of \citet{2005ApJ...627..293Y}. Then the sources with lowest luminosities are those that are observed in quiescent accretion states.

A number of methods have been applied by various authors in the past to determine distances to clouds that are either associated with large complexes or are relatively isolated \citep[e.g.,][]{1937AnHar.105..359G, 1939ApJ....89..568M, 1968AJ.....73..233R, 1969ApJ...157..611G, 1974AJ.....79...42B, 1979A&A....78..253M, 1979PASJ...31..407T, 1981ApJS...45..121S, 1986A&A...166..148R,  1990Ap&SS.166..315C, 1992BaltA...1..163C, 1992BaltA...1..149S, 1993AJ....106..672H, 1993A&A...272..235K, 1994AJ....108.1872K, 1995MNRAS.276.1052B, 1995MNRAS.276.1067B, 1997A&A...326.1215C, 1998A&A...338..897K, 1998AJ....116..881P, 2000A&A...358.1077N, 2002MNRAS.331..474F, 2004MNRAS.355.1272M, 2007A&A...470..597A, 2007ApJ...671..546L, 2010A&A...509A..44M, 2010ApJ...718..610D, 2010arXiv1006.3676K, 2010arXiv1003.2550K}. Among the methods, the trigonometric parallaxes measured using multi-epoch very long baseline array observations of radio-emitting young stars that are associated with the clouds give distances to the corresponding clouds with a precision of about 1-4\% \citep{2007ApJ...671..546L, 2008ApJ...675L..29L, 2009ApJ...698..242T}. However, this method can be applied only to those clouds that harbour radio-emitting young stars. Most of the other methods, mentioned above, are in general extremely tedious and demanding in terms of telescope time. \citet{2004MNRAS.355.1272M} in an earlier study utilized optical and Two Micron all Sky Survey (2MASS) data to determine distance to Cometary Globule, CG12. One can combine the optical data from the Naval Observatory Merged Astrometric Dataset (NOMAD) and the 2MASS data in the method given by \citet{2004MNRAS.355.1272M} to estimate distances to molecular clouds. The NOMAD database is basically a merge of data from the Hipparcos \citep{1997ESASP1200.....P}, Tycho-2 \citep{2000A&A...355L..27H}, UCAC-2 \citep{2004AJ....127.3043Z} and USNO-B1 \citep{2003AJ....125..984M} catalogues, supplemented by photometric data from the 2MASS. But the data available in this catalog is heterogeneous as the observations were carried at different epochs and also the accuracy of the photometry required by the method is high ($\leq0.1$ mag) in $BVR$. Therefore it would be highly useful if we could develop a method that utilizes the vast homogeneous $JHK_{s}$ photometric data produced by the 2MASS, already available for the entire sky, to determine distances of the clouds. The near-IR photometric method presented in a previous pilot study \citep{2010A&A...509A..44M} to estimate the cloud distances was such an endeavour. This method gives distances to clouds that are relatively closer ($\lesssim500$ pc) with a precision of $\sim20\%$.

In this paper we estimate distances to seven cloud cores namely, IRAM04191\footnote{IRAM 04191$+$1522 is the name of the protostar found associated with a cloud core not listed in the catalogue by \citet{1962ApJS....7....1L}. This core was detected in the H$^{13}$CO$^{+}$ J=1-0 survey by \citet{2002ApJ...575..950O}. The two cloud condensations identified by them towards this region are named as [OMK2002] 18a and [OMK2002] 18b. In this paper, hereafter, we call the core that harbour IRAM 04191$+$1522 as IRAM04191.}, L1521F, BHR 111, L328, L673-7, L1014, and L1148, which are identified to harbour VeLLO candidates \citep{2008ApJS..179..249D}, using near-IR photometric method \citep{2010A&A...509A..44M}. Of the seven cloud cores, two of them, IRAM04191 and L1521F are assocated with the Taurus molecular cloud complex for which very accurate distance measurements are available \citep{2007ApJ...671..546L, 2007ApJ...671.1813T, 2009ApJ...698..242T}. We included them in order to ascertain the reliability of the method. For the remaining five cores, most authors have used distances that are either guessed or assumed. The distance estimates available in the literature for each cloud core, prior to this study, are discussed along with our results in section \ref{sec:result}. This paper is organized in the following manner: a brief description of the method and the data used to estimate cloud distances is given in section \ref{sec:data}. The estimated distances of the clouds are presented in section  \ref{sec:result}. The consequences of the new distance estimates from this work on the status of the VeLLO candidates are discussed in section \ref{sec:vello}. In section \ref{sec:conclude}, we conclude the paper with a summary of our results.

\begin{table}
\caption{The J2000 coordinates of the cores as obtained from SIMBAD.}\label{tab:cord}
\begin{tabular}{lcc}\hline
Core Identification            & Right Ascension& Declination\\
                        &($^{\circ}$ $^{m}$ $^{s}$)&($^{\circ}$ $^{\prime}$ $^{\prime\prime}$)\\\hline
IRAM04191				&	04 21 56.9		&$+$15 29 47	\\ 
L1521F					&	04 30 50.3		&$+$23 00 09	\\  
BHR 111					&	15 42 20.0		&$-$52 49  06	\\    					
L328					&	18 17 00.0		&$-$18 01  54	\\
L673					&	19 20 54.4		&$+$11 13 12	\\
L1014					&	21 24 05.9		&$+$49 59 07	\\
L1148					&	20 41 10.8		&$+$67 20 35	\\
\hline
\end{tabular}
\end{table}

\section{The data and the method}\label{sec:data}
\begin{figure}
\centering
\resizebox{8.5cm}{8.5cm}{\includegraphics{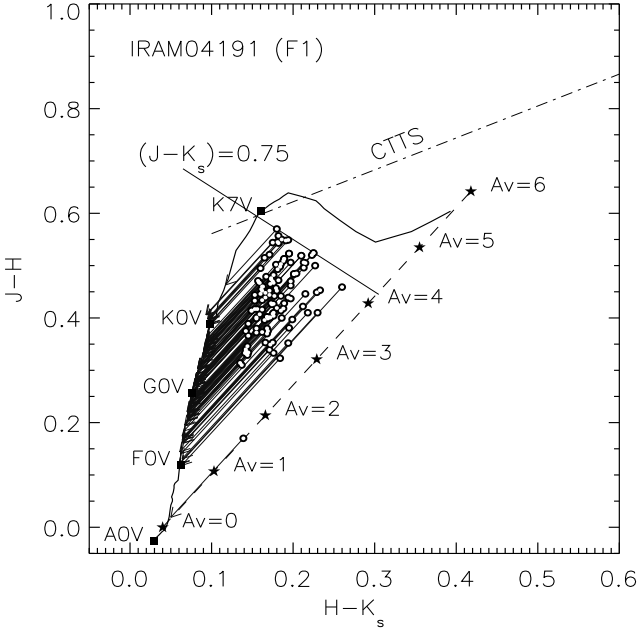}}
\caption{The ($J-H$) vs. ($H-K_{s}$) CC diagram drawn for stars (with $A_{V}\geq1$) from the region F1 towards IRAM04191 (see Fig. \ref{fig:IRAMimg}) to illustrate the method. The solid curve represents locations of  unreddened main sequence stars. The reddening vector for an A0V type star drawn  parallel to the \citet{1985ApJ...288..618R} interstellar reddening vector is shown by the dashed line. The locations of the main sequence stars of different spectral types are marked  with square symbols. The region to the right of the reddening vector is known as the NIR  excess region and corresponds to the location of PMS sources.  The dash-dot-dash line represents the loci of unreddened CTTSs \citep{1997AJ....114..288M}. The line ($J-K_{s}$)$\leq0.75$ is the upper limit set to eliminate M-type stars from the analysis as unreddened M-type stars located across the reddening vectors of A0-K7 dwarfs make it difficult to differentiate the reddened A0-K7 dwarfs from the unreddened M-type stars. The circles represent the observed colours and the arrows are drawn from the observed to the final colours obtained by the method for each star.\label{fig:CC}}
\end{figure}
Below we present a brief discussion on the method\footnote{A more rigorous discussion on the errors and limitations of the method are presented in \citet{2010A&A...509A..44M}}. We extracted $J$, $H$ and $K_{s}$ magnitudes of stars from the 2MASS all-Sky Catalog of Point Sources \citep{2003tmc..book.....C} that satisfied the following criteria, (a) photometric uncertainty\footnote{The photometric errors considered in this work  in the $J$, $H$, \& $K_{s}$ magnitudes from the 2MASS database include the corrected band photometric uncertainty, nightly photometric zero point uncertainty, and flat-fielding residual errors (Cutri et al. 2003).} $\sigma\leq0.035$ in all the three filters and (b) photometric quality flag of ``AAA'' in all the three filters, i.e., signal-to-noise ratio (SNR) $>10$. We then selected the stars with their ($J-K_{s}$)$\leq0.75$ to eliminate M-type stars from the analysis as unreddened M-type stars located across the reddening vectors of A0-K7 dwarfs make it difficult to differentiate the reddened A0-K7 dwarfs from the unreddened M-type stars. Also, the classical T Tauri stars (CTTSs), often found to be associated with the molecular clouds, occupy a well defined locus in the near infrared colour-colour (NIR-CC) diagram (Meyer et al. 1997) as shown in Fig. \ref{fig:CC} which intercepts with the main sequence loci at ($J-K_{s}$)$\approx0.75$. Thus the criterion of ($J-K_{s}$)$\leq0.75$ would allow us to eliminate most of the CTTSs as well. 

A set of dereddened colours for each star were produced from their observed colours by using a range of trial values of $A_{V}$ (0-10 mag) and the Rieke \& Lebofsky (1985) reddening law. The choice of photometric uncertainty, $\sigma\leq0.035$, ensures that the stars are selected from relatively low extinction regions of the cloud where the Rieke \& Lebofsky (1985) reddening law is most applicable \citep{1982Ap&SS..85..271S, 2003AJ....126.1888K}. The computed sets of dereddened colours of a star were then compared with the intrinsic colours of the normal main sequence stars. The intrinsic colours of the main sequence stars in the spectral range  A0-K7 were taken from \citet{2010A&A...509A..44M}. The best match giving a minimum value of $\chi^{2}$ then yielded the spectral type and $A_{V}$ corresponding to that intrinsic colour. Once the spectral types and $A_{V}$ values of the stars were known, their distances were estimated using the distance equation, $d~(pc) = 10^{(K_{s}-M_{K}+5-A_{K})/5}$. Only those stars that were classified as dwarfs were considered for the determination of distances as the absolute magnitudes of giants are highly uncertain. The whole procedure is illustrated in Fig.\ref{fig:CC} where we plot the NIR-CC diagram for stars (with $A_{V}\geq1$) chosen from the region F1 towards the direction of IRAM04191 (see Fig. \ref{fig:IRAMimg}). The arrows are drawn from the  observed data points to the corresponding dereddened colours estimated using the method. The maximum extinction values that can be measured using the method are those for A0V type stars ($\approx4$ magnitude).  The extinction traced by stars will fall as we move towards more late type stars.

\begin{figure}
\centering
\resizebox{8cm}{12cm}{\includegraphics{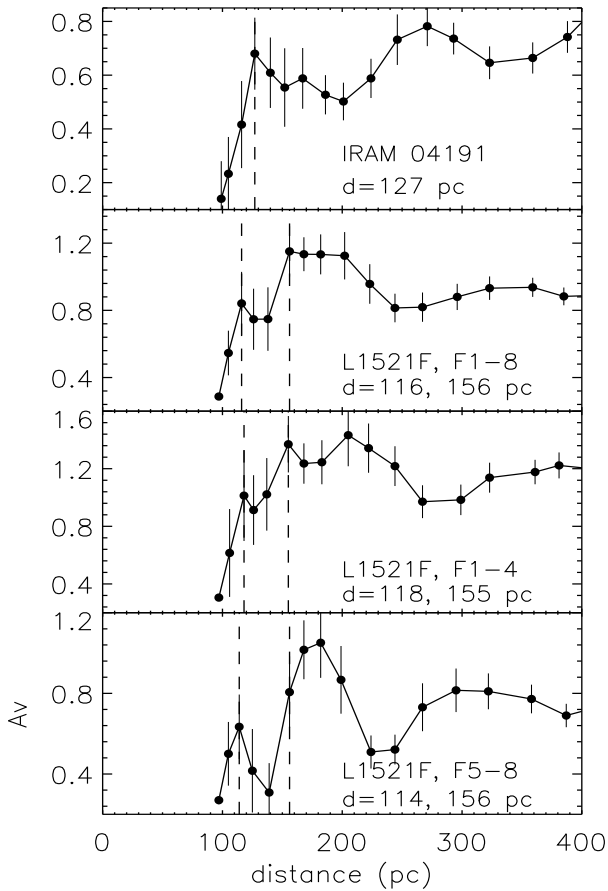}}
\caption{The mean values of $A_{V}$ vs. the mean values of distance plot for  IRAM04191 and L1521F produced using the the procedure discussed in the \S\ref{sec:data} to determine distances to the clouds. The error bars on the mean $A_{V}$ values were calculated using the expression, $standard~deviation/\sqrt[]{N}$, where $N$ is the number of stars in each bin.\label{fig:taurus_hist}}
\end{figure}

\begin{figure}
\centering
\resizebox{8cm}{15cm}{\includegraphics{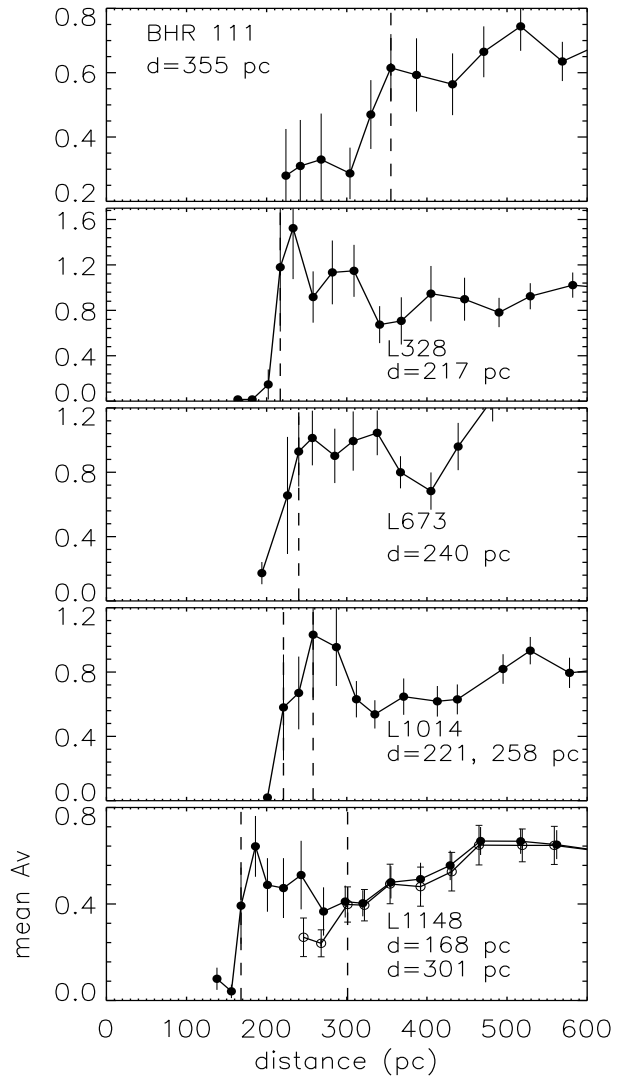}}
\caption{ Same as in Fig. \ref{fig:taurus_hist} but for BHR 111, L328, L673, L1014 and L1148.\label{fig:hist}}
\end{figure}

In order to estimate distance to a cloud (not just by eye estimation), we first grouped the stars, classified as dwarfs, into distance bins of $bin~width = 0.18\times distance$. The centers of each bin were kept at a separation of half of the bin width. Because there exist very few stars at smaller distances, the mean value of the distances and the $A_{V}$ of the stars in each bin were calculated by taking 1000 pc as the initial point and proceeded towards smaller distances. The mean distance of the stars in the bin at which a  significant drop in the mean of the extinction occurred was taken as the distance to the cloud and the average of the uncertainty in the distances of the stars in that bin was taken as the final uncertainty of the determined distance of the cloud (see Figs. \ref{fig:taurus_hist} and \ref{fig:hist}). The vertical dashed line in the $A_{V}$ vs. $d$ plots, used to mark the cloud distance, is drawn at a distance deduced from the above procedure. The error in the mean values of $A_{V}$ were calculated using the expression, $standard~deviation/\sqrt[]{N}$, where $N$ is the number of stars in each bin. The $A_{V}$ values were found to be uncertain by $\sim0.6$ magnitude. The typical errors in our distance estimates for the clouds were found to be $\sim18\%$ \citep{2010A&A...509A..44M}.

One of the prominent features of the molecular cloud structure is that it is hierarchical. The extinction maps produced towards a number of regions in our Galaxy, using optical and near infrared techniques, have shown that the high-density features are invariably contained inside an extended low-density envelope \citep[e.g.,][]{2005PASJ...57S...1D, 2010A&A...512A..67L}. Our method utilizes the background stars shining through the outer low-extinction ($A_{V}\lesssim4$ magnitude) regions of the cores to estimate their distances. In case of the presence of a single dust layer in the line of sight, we expect a sharp increase in the extinction for the stars that are located behind that dust layer in an $A_{V}$ vs. $d$ plot. The distance at which the sharp increase in the extinction occurs is considered as the distance to that dust layer. But if there exist more than one dust layer along the line of sight, we expect steep but step-like increase in the extinction at distances at which the dust layers are located. This, however, happens only if the foreground dust layers contribute a constant extinction values. In case of patchy foreground dust layer(s), this method can give distance only to the first layer as the sharp step-like features get smeared by the non-uniform extinction caused by the foreground dust layer.  

In a recent work, \cite{2010arXiv1006.3676K}, using the 2MASS data, has presented a similar method to estimate distances to molecular clouds. Instead of using the main sequence stars in the spectral range from A0-K7 as in our work, \cite{2010arXiv1006.3676K} utilized three spectral groups: O-G6 as primary sources, M4-T as secondary sources and G6-M0 as teritary sources in the estimation of the cloud distances. The absolute magnitude in J-band were calibrated using Hipparcos stars. Following a rigorous statistical approach, the jump in the extinction in the A$_{V}$ vs. distance plot was obtained to determine the distance. In the pilot study, \cite{2010arXiv1006.3676K} estimated distances to a number of nearby clouds that includes Chamaeleon I and Lupus 3 which were included in \cite{2010A&A...509A..44M} also. The distance estimates in both the works are found to be in good agreement within the error (the typical error in the distance estimates of \cite{2010arXiv1006.3676K} is found to be $\sim8\%$). 
\begin{table}
\centering
\caption{The details of the fields selected towards each cloud.}\label{tab:details}
\begin{tabular}{lccclr}\hline
\#&Area           &$l$              &$b$            &Total &Stars classified\\
     &selected&&&stars &as Dwarfs\\
     &($^{\circ}$)&($^{\circ}$)  &($^{\circ}$) &&\\\hline
          \multicolumn{4}{l}{IRAM04191}\\
1	&1.5               &178.7968	&$-$23.9180	&403	&277	\\
2	&1.0               &180.2393	&$-$23.9180	&438	&270	\\\hline
    &                    &                & Total     	&841	&547	\\\hline
              \multicolumn{4}{l}{L1521F}\\
1&  1.0&172.5343  &$-$13.5645&199&155\\
2&  1.0&171.5169  &$-$13.5645&444&331\\
3&  1.0&171.5170  &$-$14.5545&160&120\\
4&  1.0&172.5388  &$-$14.5545&144&97\\
5&  1.0&171.8592  &$-$15.5445&368&243\\
6&  1.0&170.8327  &$-$15.5445&285&197\\
7&  1.0&170.8294  &$-$16.5345&558&386\\
8&  1.0&171.8610  &$-$16.5345&438&280\\\hline
    &                    &                & Total     &2596  &1809\\\hline
         \multicolumn{4}{l}{BHR111}\\
 1&0.5&327.3478  & $+$1.9000&327&271\\
 2&0.5&327.1521  & $+$1.4000&286&250\\
 3&0.5&327.2173  & $+$0.9000&265&220\\\hline
    &                    &                & Total     & 878&741\\\hline
          \multicolumn{4}{l}{L328}\\
 1& 0.3&12.9344 &  $-$0.8481&74 &62\\
 2& 0.5&12.5573 &  $-$0.5035&328&285\\\hline
    &                    &                & Total     & 402 &347\\\hline
         \multicolumn{4}{l}{L673-7}\\
1& 0.5& 46.1434 &  $-$1.3152&34 &22\\
2& 0.5& 46.4786 &  $-$0.8260&108&88\\
3& 0.5& 45.9856 &  $-$0.8260&111&89\\\hline
    &                    &                & Total     & 253 &199\\\hline
         \multicolumn{4}{l}{L1014}\\
1& 0.4& 92.3377 &  $-$0.2672 &253&209\\
2& 0.4& 92.7318 &  $-$0.0560 &229&191\\\hline
    &                    &                & Total     & 482 &400\\\hline
         \multicolumn{4}{l}{L1148}\\
1 &0.7&102.5346   &$+$15.2936&191&144\\
2 &0.7&101.7809   &$+$15.1666&205&151\\
3 &0.7&102.5360   &$+$16.0555&196&145\\
4 &0.7&101.7795   &$+$15.9285&208&149\\\hline
    &                    &                & Total     & 800 &589\\\hline
\hline
\end{tabular}
\end{table}
\section{Results and Discussion}

\subsection{Distances to the cloud cores IRAM04191, L1521F, BHR111, L328, L673-7, L1014, and L1148\label{sec:result}}

We applied the method on seven dense cores, namely, IRAM04191, L1521F, BHR111, L328, L673-7, L1014, and L1148. We divided the fields containing the cores into small  sub-fields to avoid complications created by the erroneous classifications of giants into dwarfs. While the rise in the extinction due to the presence of a cloud should occur almost at the same distance in all  the fields, if the whole cloud is  located at the same distance, the wrongly classified stars in the sub-fields would show high extinction not at the same but at random  distances \citep{2010A&A...509A..44M}. In the case of cores that have small angular sizes, we included fields containing additional cores that are located spatially closer and show similar radial velocities (obtained, for example, from CO observations), in order to have sufficient number of stars to infer their distances. Here we assume that the cores that are spatially closer and have similar velocities are located at almost the same distances. Table \ref{tab:details} summarizes the area of the fields (in degree) chosen towards the cores studied here, the galactic coordinates of the center of the fields,  the  total number of stars selected after applying all  the selection  criteria ($\sigma\leq0.035$, $SNR>10$, \& ($J-K_{s}$)$\leq0.75$) and the number of stars classified as dwarfs by our method towards each field. Below we present results and discussion on individual clouds. In the $A_{V}$ vs. $d$ plots presented for the cores studied, the dash-dotted curve shows the increase in the extinction towards the clouds' galactic latitude  as a function of distance  produced using the expressions given by \citet{1980ApJS...44...73B} (BS80, hereafter). The dashed vertical line(s) is drawn at the cloud distance(s) inferred from the procedure described in \S \ref{sec:data} (see Figs. \ref{fig:taurus_hist} and \ref{fig:hist}).

\subsubsection{IRAM04191}
\begin{figure}
\centering
\resizebox{9cm}{7cm}{\includegraphics{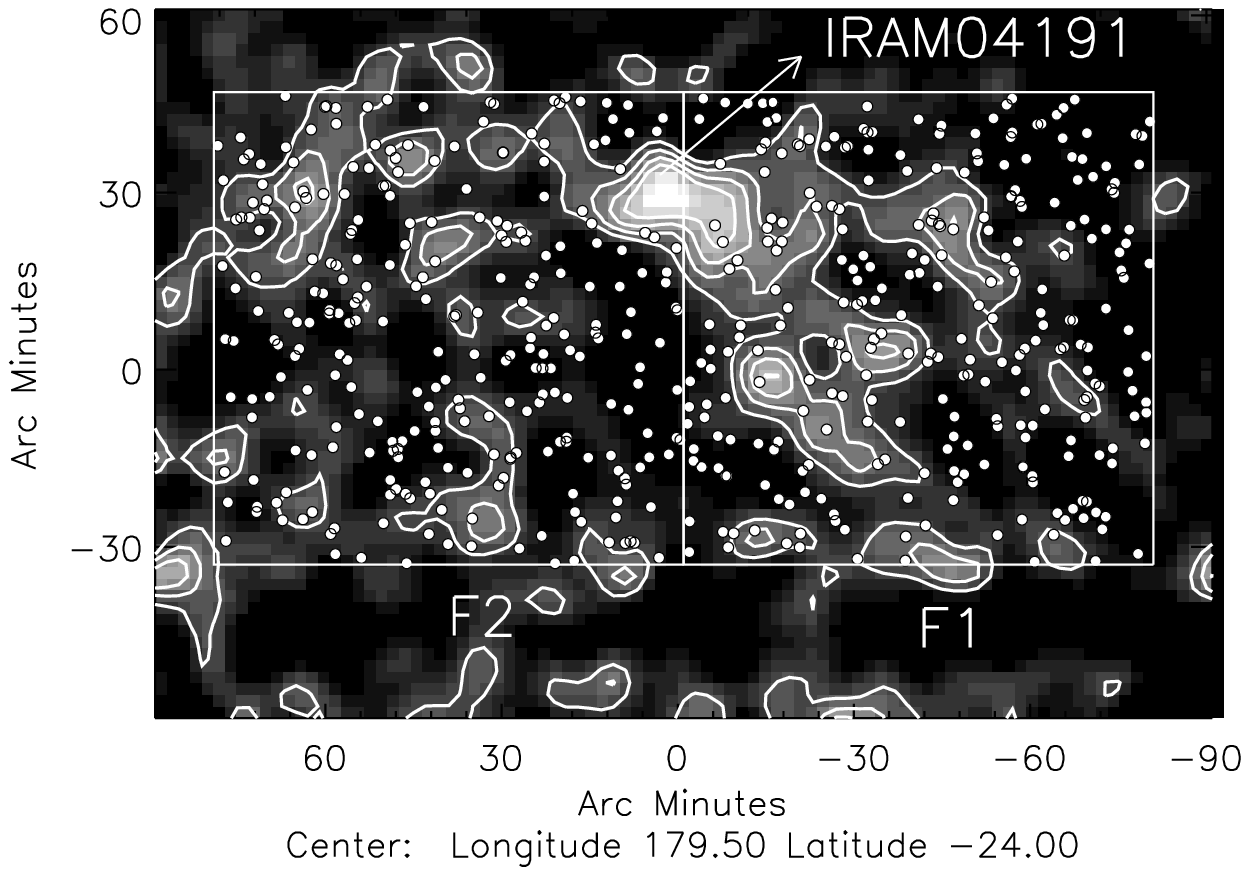}}
\caption{The $2^{\circ}\times3^{\circ}$ extinction map produced by \citet{2005PASJ...57S...1D} of the region containing IRAM04191. The contours are drawn at 0.4, 0.6, 0.8 and 1.0 magnitude levels. The fields used for selecting star to determine distance are identified and labelled. The stars classified as dwarfs and used for determining distance to the cloud are identified using open circles.\label{fig:IRAMimg}}
\end{figure}
\begin{figure}
\centering
\resizebox{9cm}{6.5cm}{\includegraphics{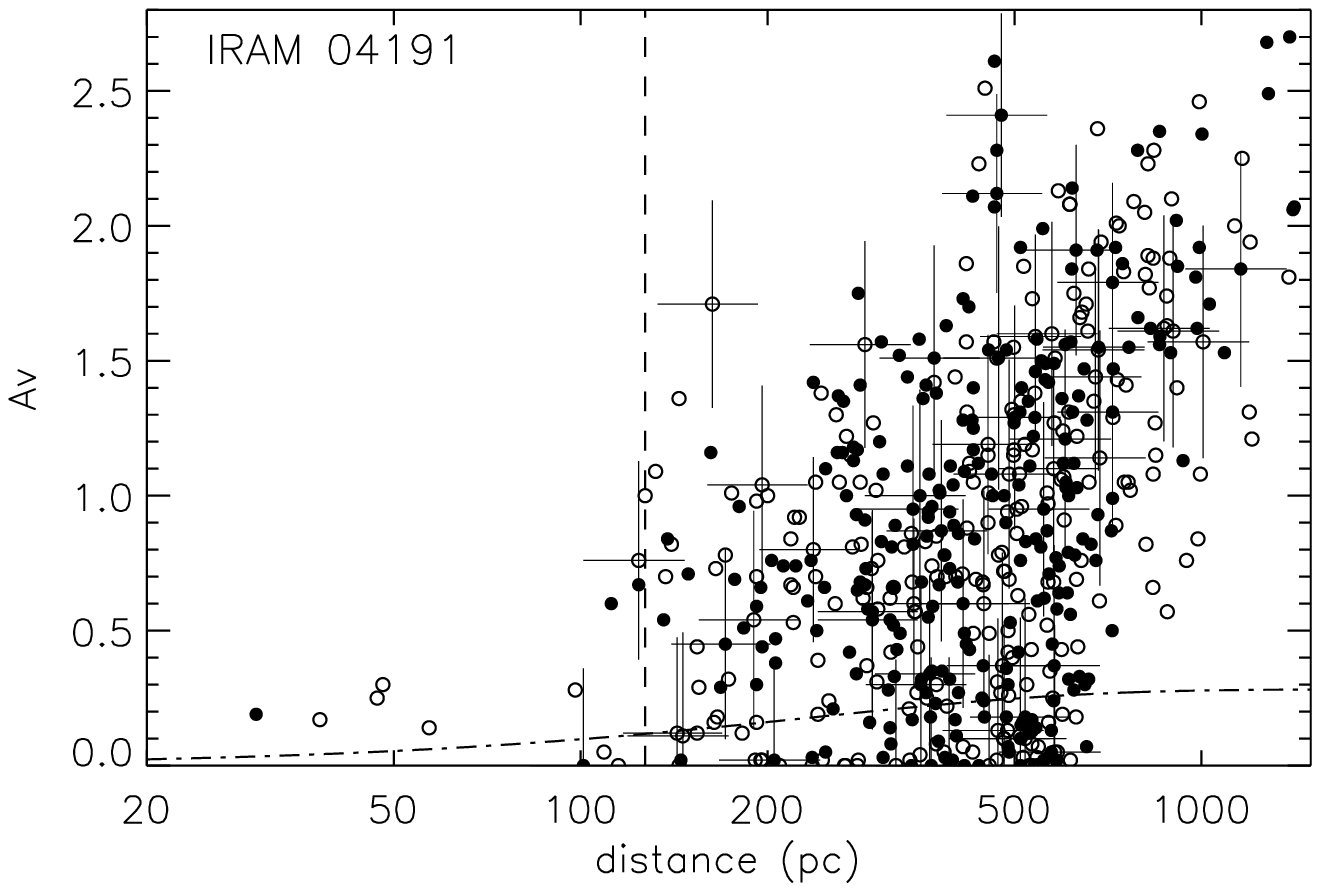}}
\caption{The $A_{V}$ vs. $d$ plot for all  the stars classified as dwarfs from the fields F1-2 combined together towards IRAM04191. The dashed vertical line is drawn at 127 pc inferred from the procedure described in the \S \ref{sec:data} (see Fig.\ref{fig:hist}). The dash-dotted curve represents  the increase in the extinction towards the Galactic latitude of $b=-24^{\circ}$ as a function of distance  produced from the expressions given by BS80. The error bars are shown on a few stars for better clarity of the points.\label{fig:allIRAMdist}}
\end{figure}

We show the $3^{\circ}\times2^{\circ}$ extinction map~\footnote{The extinction map, covering the entire region in the galactic latitude range $|b|\leq40^{\circ}$ derived using the optical database ``Digitized Sky Survey I'' and the traditional star-count technique, was produced in two angular resolutions of $6^{\prime}$ and $18^{\prime}$ (Dobashi et al. 2005). In this paper, we used the maps with $6^{\prime}$ angular resolution.} produced by \citet{2005PASJ...57S...1D} of the region containing IRAM04191 in Fig. \ref{fig:IRAMimg}. The contours are drawn at 0.4, 0.6, 0.8 and 1.0 magnitude levels. We selected two fields, F 1 and F2, each covering an area of $1^{\circ}\times1^{\circ} $ as shown in Fig. \ref{fig:IRAMimg}. The stars, classified as dwarfs, that are used for determining the distance to the cloud are identified using circles.

In Fig. \ref{fig:allIRAMdist} we present the $A_{V}$ vs. $d$ plot for the stars from both F1 (filled circles) and F2 (open circles) combined together. The dashed vertical line is drawn at 127 pc. The jump in the extinction values significantly above of that expected from the expression of BS80 is found to occur, in both fields, at $\sim127$ pc. Beyond this distance, the stars with high extinction are found to be distributed almost continuously in distance as one would expect since the cloud will act as a dense sheet of dust layer dimming the stars shining through it. Using 541 stars that are classified as main sequence stars, we estimated a distance of $127\pm25$ pc to the cloud IRAM04191.

Very accurate distance determinations ($\lesssim1\%$) are available for  a number of naked T Tauri stars associated with the Taurus molecular cloud complex from the trigonometric parallax measurements made using the Very Long Baseline Array (VLBA) multi-epoch observations carried out by detecting their non-thermal 3.6 cm radio continuum emission. These stars include, HDE 283572, Hubble 4, T Tau and HP Tau/G2. They are found to be located at distances of $128.5\pm0.6$ pc \citep{2007ApJ...671.1813T}, $ 132.8\pm0.5 $ pc \citep{2007ApJ...671.1813T}, $ 147.6\pm0.6 $ pc \citep{2007ApJ...671..546L} and $ 161.2\pm0.9 $ pc \citep{2009ApJ...698..242T} respectively. As of now, from the distances determined for these four stars, it is apparent that the clouds in the Taurus molecular cloud complex are distributed in space in distances ranging from $\sim128$ pc to $\sim161$ pc. From the mean parallax obtained from the above four stars \citet{2009ApJ...698..242T} determined a mean distance of $141.2$ pc to the Taurus cloud complex. At a distance of $127\pm25$ pc, IRAM04191 could possibly located at the front end of the Taurus cloud complex. In all the previous studies of IRAM04191, this cloud was considered to be associated with the Taurus cloud complex and assigned a distance of 140 pc \citep{1994AJ....108.1872K}. The distance estimate presented here is the first independent confirmation of the association of IRAM04191 with the Taurus cloud complex.
\subsubsection{L1521F}

\begin{figure}
\centering
\resizebox{9cm}{11cm}{\includegraphics{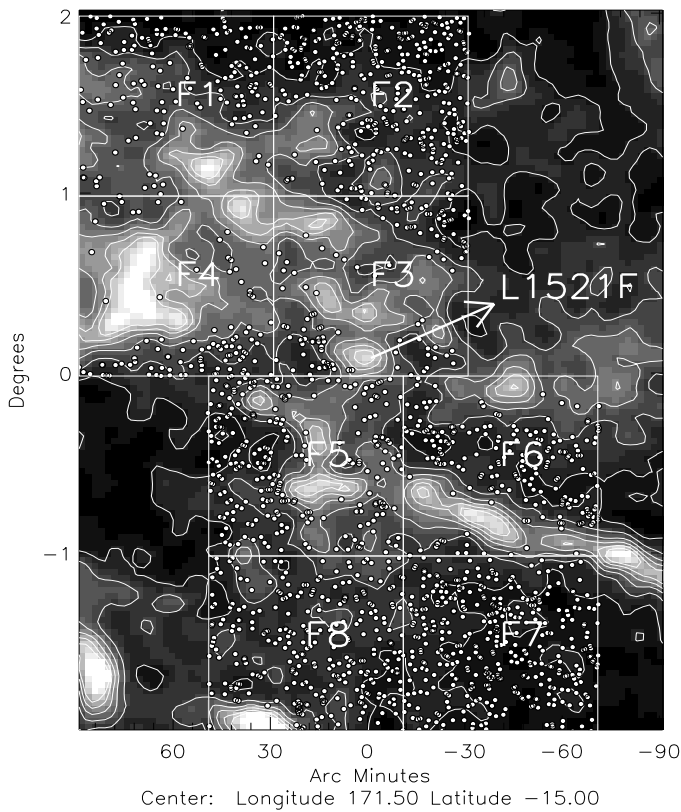}}
\caption{The $\sim3^{\circ}\times4^{\circ}$ extinction map produced by \citet{2005PASJ...57S...1D} of the region containing L1521F. The contours are drawn at  0.5, 1.0, 1.5, 2.0, 2.5 and 3.0 magnitude levels. The fields used for selecting star to determine distance are identified and labelled. L1521F is located in F3.\label{fig:1521img}}
\end{figure}
\begin{figure}
\centering
\resizebox{9cm}{6.5cm}{\includegraphics{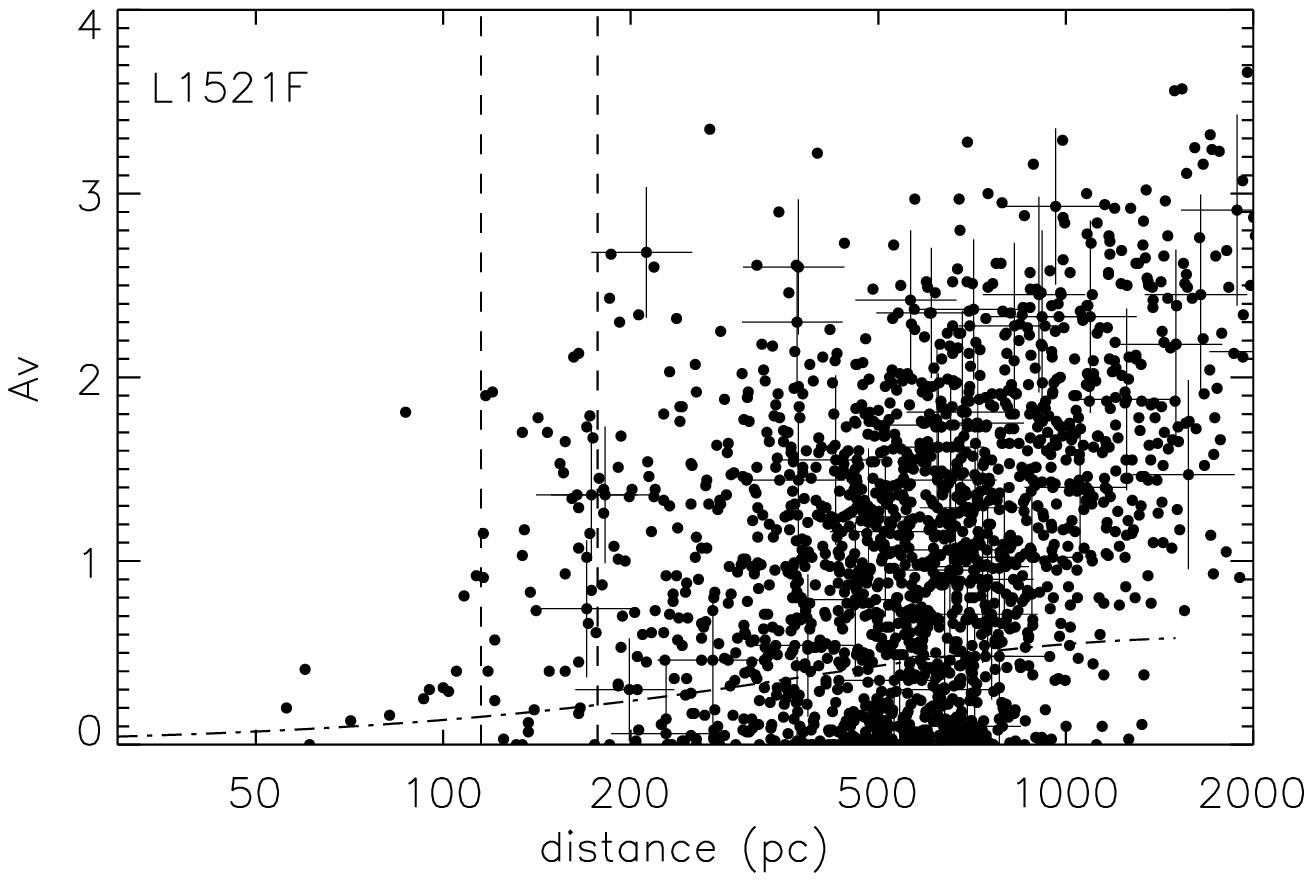}}
\caption{The $A_{V}$ vs. $d$ plot for all   the stars classified as dwarfs from the fields F1-8 combined together towards L1521F. The dashed vertical lines are drawn at distances 116 pc and 156 pc inferred from the procedure described in the \S \ref{sec:data} (see Fig.\ref{fig:hist}). The dash-dotted curve represents  the increase in the extinction towards the Galactic latitude of $b=-14.9^{\circ}$ as a function of distance  produced from the expressions given by BS80. The error bars are shown on a few stars for better clarity of the points.\label{fig:all1521dist}}
\end{figure}
\begin{figure}
\centering
\resizebox{9cm}{12cm}{\includegraphics{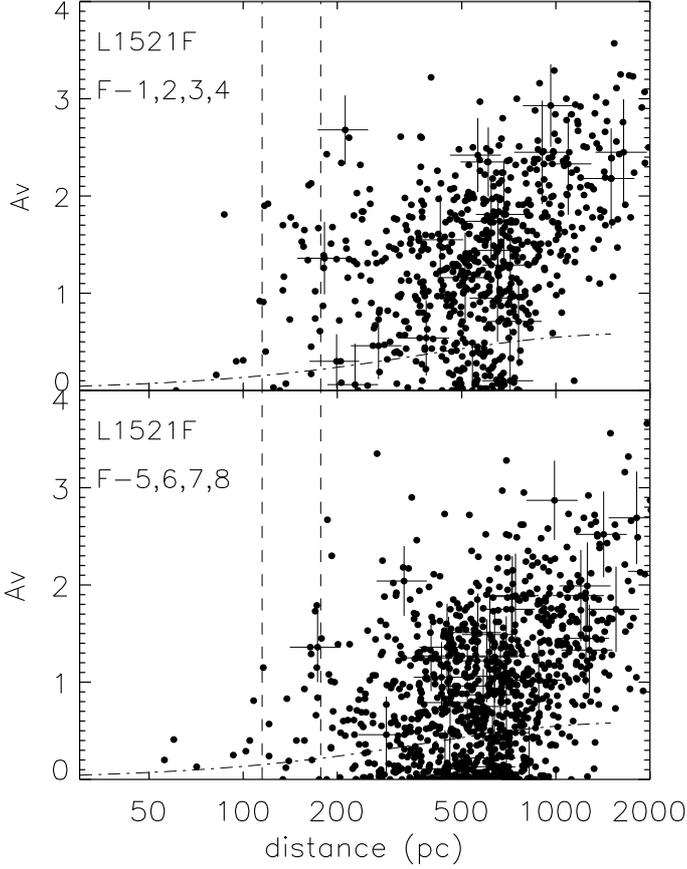}}
\caption{The $A_{V}$ vs. $d$ plot for the stars classified as dwarfs from the fields F1-4 in the upper panel and F5-8 in the lower panel towards L1521F. The dashed vertical lines are drawn at 116 pc and 156 pc. The dash-dotted curve represents  the increase in the extinction towards the Galactic latitude of $b=-14.9^{\circ}$ as a function of distance  produced from the expressions given by BS80. The error bars are shown on a few stars for better clarity of the points.\label{fig:1521dist}}
\end{figure}

Figure \ref{fig:1521img} shows the $\sim3^{\circ}\times4^{\circ}$ extinction map of the region containing L1521F. The contours are drawn at 0.5, 1.0, 1.5, 2.0, 2.5 and 3.0 magnitude levels. We divided the region containing L1521F into eight fields, F1-8, as shown in Fig. \ref{fig:1521img}. Each field covers an area of  $1^{\circ}\times1^{\circ} $. The stars, classified as dwarfs, that are used for determining distance to the cloud are identified using circles. The cloud L1521F is located in the field F3.

In Fig. \ref{fig:all1521dist}, we present $A_{V}$ vs. $d$ plot for the stars from all  the fields (F1-8) combined together.  The dashed vertical lines are drawn at 116 and 156 pc. Though, in Fig. \ref{fig:all1521dist}, only one component of dust layer is apparent, we could notice the presence of two dust layers in Fig. \ref{fig:taurus_hist} as it shows a sharp rise in the mean values of extinction occurring at two distances, one at 116 pc and another at 156 pc. In the $A_{V}$ vs. $d$ plots of individual fields, we found that the presence of a dust layer at 116 pc is more conspicuous towards the fields F1-4,  than towards the fields F5-8. In Fig. \ref{fig:1521dist}, we present the $A_{V}$ vs. $d$ plot for the stars from the fields F1-4 (upper panel) and F5-8 (lower panel) separately to show the dominance of the two components of dust layers in these two regions. In Fig. \ref{fig:taurus_hist} we show the distances determined using the stars from F1-4 and F5-8 independently. The two dust layers are found to be at distances $118\pm22$ and $155\pm29$ pc and $114\pm21$ and $156\pm29$ pc towards the fields F1-4 and F5-8 respectively. The distances estimated for the two dust components using stars from separate sets of  fields are found to be in good agreement. These results ascertain the fact that the technique employed here in estimating distances to dust layers in a direction is capable of picking them consistently even though the fields ($1^{\circ}\times1^{\circ} $ each) used for selecting the stars are located relatively far (3-4$^{\circ}$ in this case) apart. 

The dust layer at $\sim156$ pc seems to be present in all the eight fields. Also,  the $A_{V}$ vs. $d$ plot produced for the sources from F5-8 do contain a number of stars that show high extinction at distances smaller than $\sim156$ pc. This could be due to the contribution from the dust component at $\sim114$ pc. One of the limitations of the method is that in case of multiple jumps in extinction in an $A_{V}$ vs. $d$ plot, it is difficult to associate a particular jump with any of the dust layer. In fact, this is true for all methods that utilize the rise in the extinction as a way to estimate distances.

The above findings encouraged us to estimate distances to the entire Taurus molecular cloud complex. We found the presence of four dust components at distances $85\pm16$, $114\pm20$, $ 138\pm25 $  and $168\pm31$ pc towards it. The results will be presented in a forthcoming paper. The presence of the dust component at $\sim85$ pc, though, is found to be less evident towards the entire cloud complex. Based on the fact that the dust layer at $\sim118$ pc is found to be dominant towards F3, the field in which L1521F is located, we associate the jump at $\sim118\pm22$ pc to the cloud L1521F.  But adopting a mean value of $136\pm36$ pc (error in this case is $\sqrt{21^{2}+29^{2}}$) to L1521F would be more appropriate as it is closer to the distances of two sources, HDE 283572 and Hubble 4, found to be located nearer ($<2^{\circ}$) to L1521F and are at distances  $128.5\pm0.6$ pc and $ 132.8\pm0.5 $ pc respectively \citep{2009ApJ...698..242T}. 
\subsubsection{BHR 111}
\begin{figure}
\centering
\resizebox{9cm}{11cm}{\includegraphics{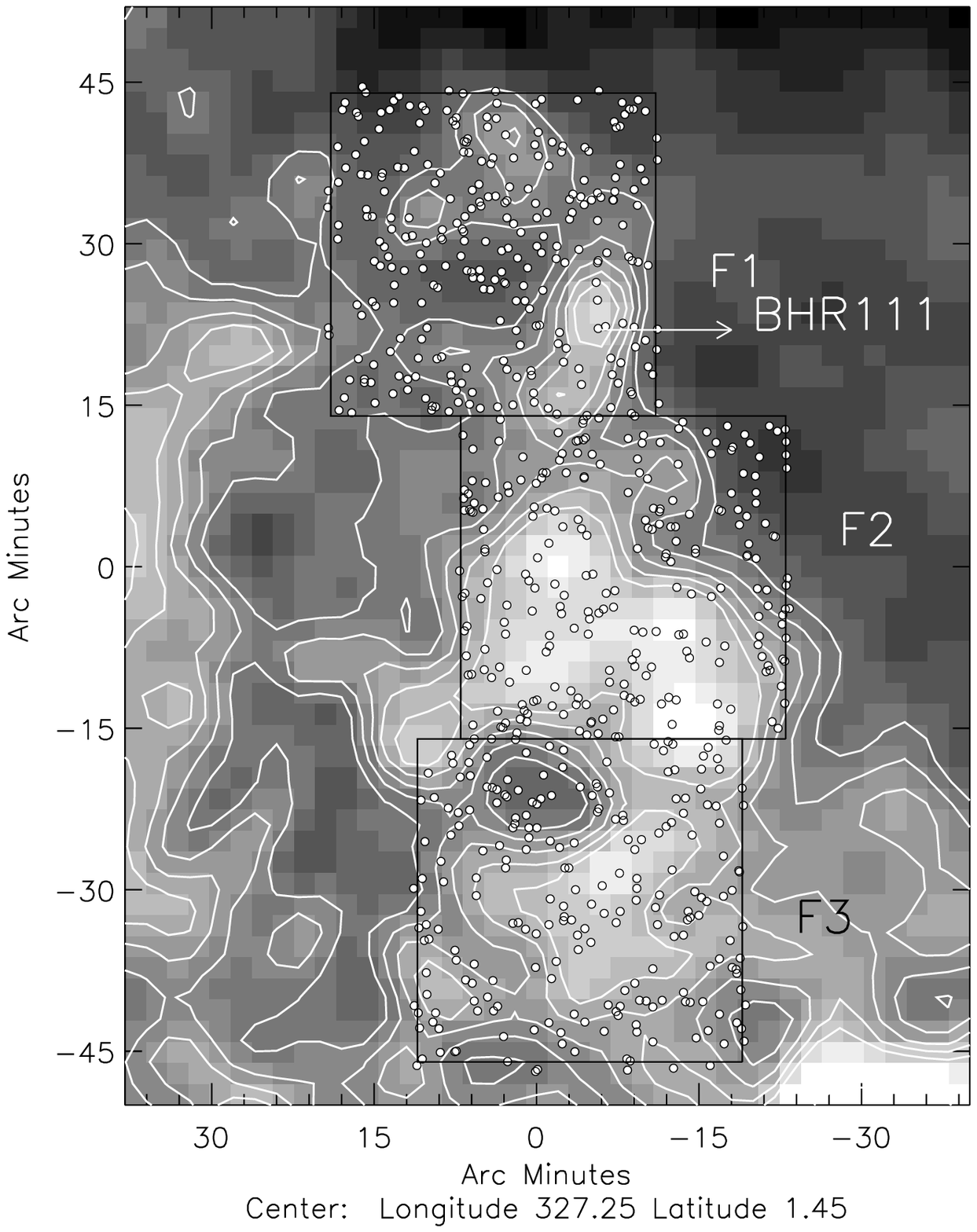}}
\caption{The $\sim1.5^{\circ}\times1.5^{\circ}$ extinction map produced by \citet{2005PASJ...57S...1D} of the region containing BHR 111. The contours are drawn at 2, 2.1, 2.2, 2.3 and 2.4 magnitude levels. The fields used for selecting star to determine distance are identified and labelled. The stars classified as dwarfs and used for determining distance to the cloud are identified using open circles.\label{fig:BHRimg}}
\end{figure}
\begin{figure}
\centering
\resizebox{9cm}{6.5cm}{\includegraphics{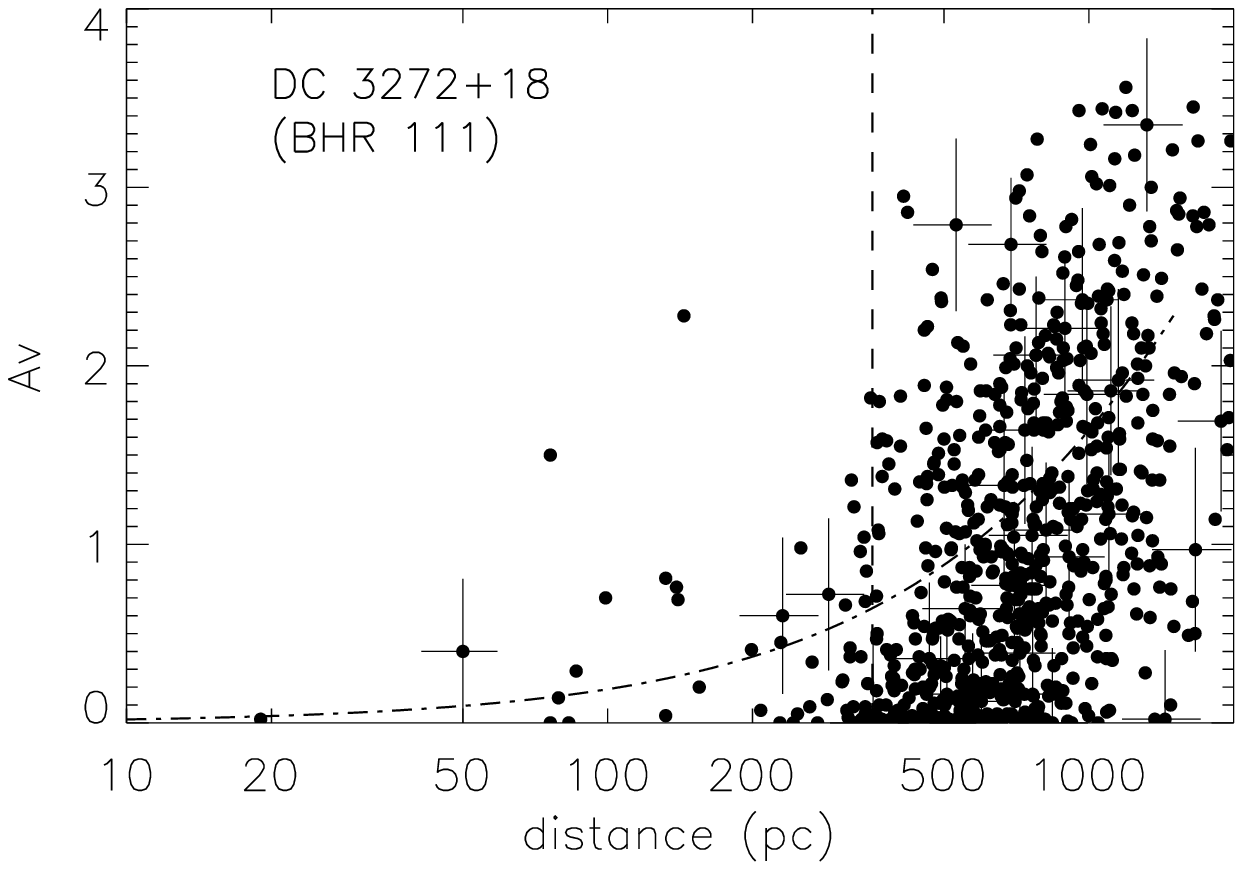}}
\caption{The $A_{V}$ vs. $d$ plot for all  the stars classified as dwarfs from the fields F1-3 combined together towards BHR 111. The dashed vertical line is drawn at 355 pc inferred from the procedure described in the \S \ref{sec:data} (see Fig.\ref{fig:hist}). The dash-dotted curve represents  the increase in the extinction towards the Galactic latitude of $b=1.824^{\circ}$ as a function of distance  produced from the expressions given by BS80. The error bars are shown on a few stars for better clarity of the points.\label{fig:allBHRdist}}
\end{figure}
\begin{figure}
\centering
\resizebox{9cm}{13cm}{\includegraphics{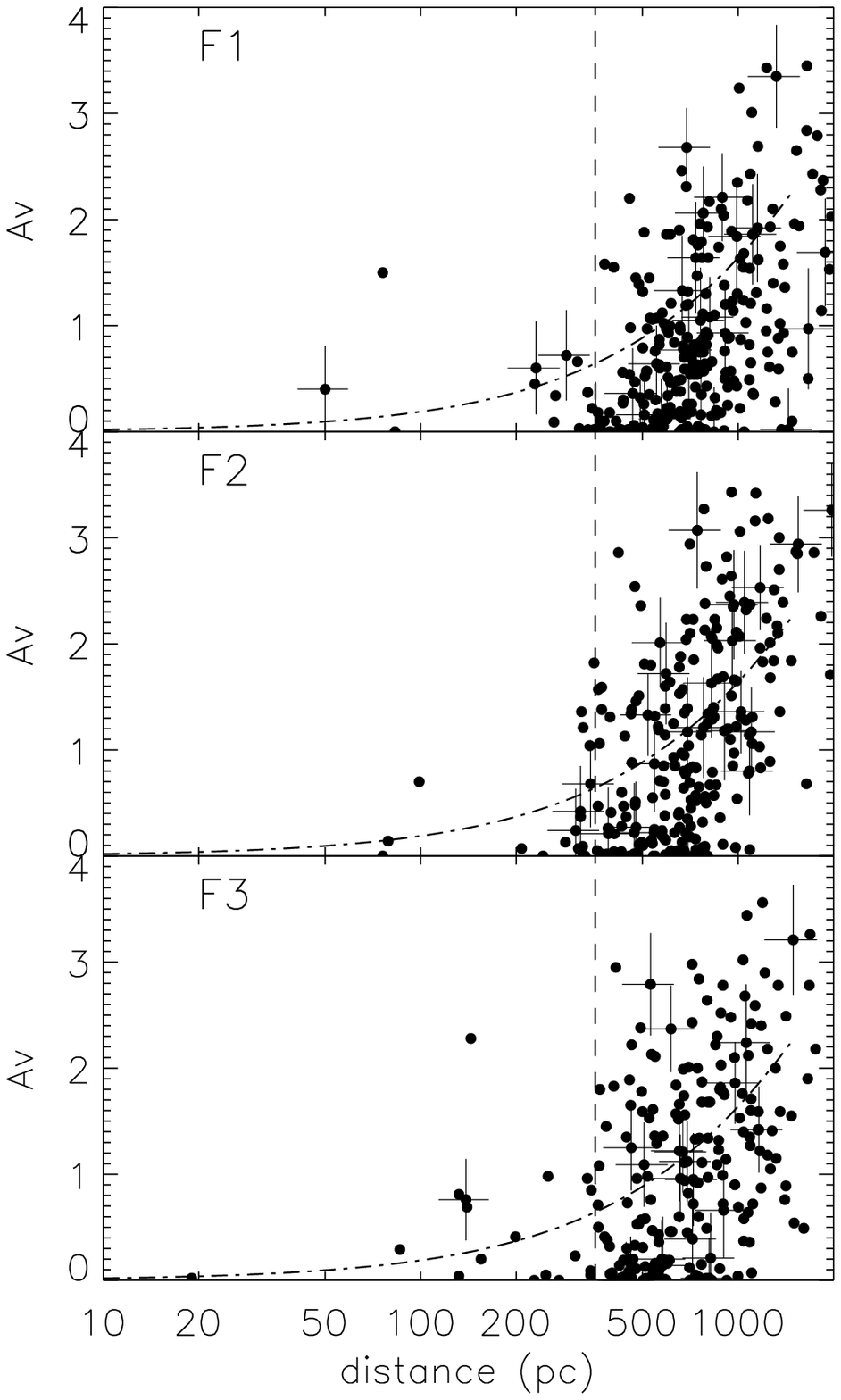}}
\caption{The $A_{V}$ vs. $d$ plot for the stars classified as dwarfs from the individual fields F1, F2, \& F3 towards BHR 111. The dashed vertical line is drawn at 355 pc inferred from the procedure described in the \S \ref{sec:data} (see Fig.\ref{fig:hist}). The dash-dotted curve represents  the increase in the extinction towards the Galactic latitude of $b=1.824^{\circ}$ as a function of distance  produced from the expressions given by BS80. The error bars are shown on a few stars for better clarity of the points.\label{fig:BHRdist}}
\end{figure}

In Fig. \ref{fig:BHRimg}, we present the $\sim1.5^{\circ}\times1.5^{\circ}$ extinction map of the region containing BHR 111 (DC 3272+18). The contours are drawn at 2, 2.1, 2.2, 2.3 and 2.4 magnitude levels. The size of the cloud is only $10^{\prime}\times4^{\prime}$ \citep{1995MNRAS.276.1052B}. We included two additional fields F2 and F3 apart from the field F1, which contains BHR 111, as shown in Fig. \ref{fig:BHRimg}. Unfortunately, no velocity information is available on the regions selected except for BHR 111 \citep{1995MNRAS.276.1052B}. The stars, classified as dwarfs, that are used for determining the cloud distance are identified using circles.

In Fig. \ref{fig:allBHRdist} we present the $A_{V}$ vs. $d$ plot for the stars from all  the fields (F1-3) combined together. The dashed vertical line is drawn at 355 pc. There are a number of stars that exhibit relatively high extinction values at distances smaller than 355 pc. Inspection of Fig. \ref{fig:BHRdist}, which presents the $A_{V}$ vs. $d$ plot for stars from the individual fields separately, shows that the jump in the extinction values consistently occurs at or beyond 355 pc. The presence of a foreground dust layer is suspected. On the basis of 741 stars classified as dwarfs by the method, we estimate a distance of $355\pm65$ pc to BHR 111.

Distance to BHR 111 was first determined by \citet{1995MNRAS.276.1067B} from a plot of stellar reddening of stars, selected from a circular region about the object, as against their distances corrected for reddening. They selected stars that had MK spectral types and colours available. They estimated a distance of 250 pc to BHR 111. Using the same method, \citet{2009ApJ...703.1444R} also estimated a distance of 250 pc to BHR 111. \citet{1995MNRAS.276.1067B} used a search radius of $5^{\circ}$ about the cloud to obtain stars with MK spectral types and colours. If not enough stars were available within $5^{\circ}$, they increased the search radius to $7.5^{\circ}$ or $10^{\circ}$.  \citet{2009ApJ...703.1444R} used a search radius of $3^{\circ}$ about the cloud to select the stars. In Fig. \ref{fig:BHRimg}, it can be noticed that the cloud is not isolated but is surrounded by diffused dust clouds. In Fig. \ref{fig:BHRdist} we noticed the evidence of more intervening dust layers towards the line-of-sight which is more conspicuous towards F3. It could be possible that the large search radius used by both \citet{1995MNRAS.276.1067B} and \citet{2009ApJ...703.1444R} might have picked the nearer dust layer possibly located at 250 pc.

\subsubsection{L328}\label{res:l328}

\begin{figure}
\centering
\resizebox{9cm}{8cm}{\includegraphics{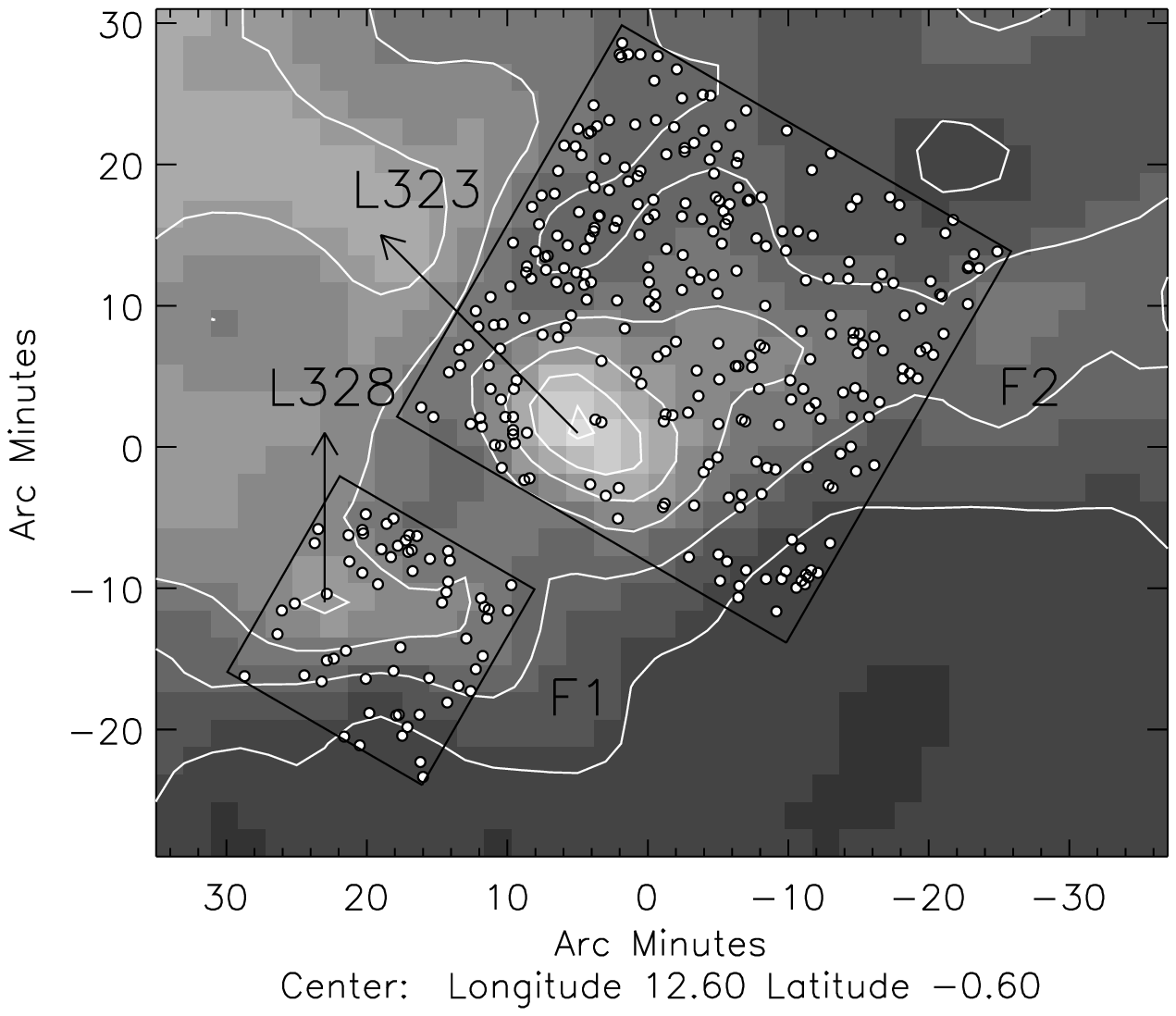}}
\caption{The $\sim1^{\circ}\times1^{\circ}$ extinction map produced by \citet{2005PASJ...57S...1D} of the region containing L328. The contours are drawn at 1.5, 2.0, 2.5, 3.0, 3.5 and 4.0 magnitude levels. The fields used for selecting star to determine distance are identified and labelled. The stars classified as dwarfs using the near-IR photometry in each field are shown using filled circles. The fields F1 and F2 cover an area of  $ 0.3^{\circ}\times0.3^{\circ} $ and $ 0.5^{\circ}\times0.5^{\circ} $ respectively.\label{fig:328img}}
\end{figure}
\begin{figure}
\centering
\resizebox{9cm}{6.5cm}{\includegraphics{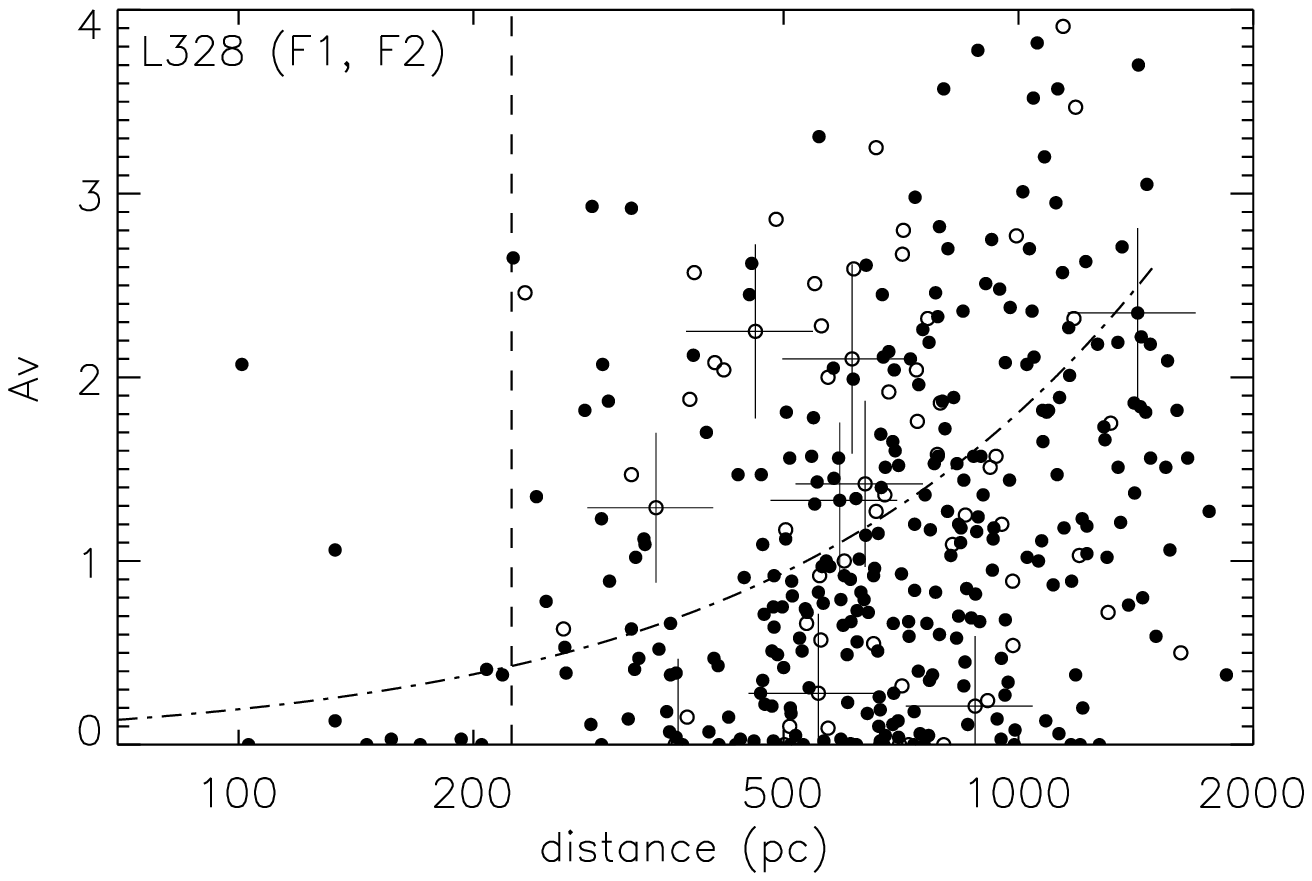}}
\caption{The $A_{V}$ vs. $d$ plot for all  the stars classified as dwarfs from the fields F1 (open circles) \& F2 (filled circles) combined together towards L328. The dashed vertical line is drawn at 217 pc inferred from the procedure described in the \S \ref{sec:data} (see Fig.\ref{fig:hist}). The dash-dotted curve represents  the increase in the extinction towards the Galactic latitude of $b=-0.829^{\circ}$ as a function of distance  produced from the expressions given by \citet{1980ApJS...44...73B}. The error bars are shown on a few stars for better clarity of the points.\label{fig:328alldist}}
\end{figure}

In Fig. \ref{fig:328img} we present the $1^{\circ}\times1^{\circ}$ extinction map of the region containing L328. The contours are drawn at 1.5, 2.0, 2.5, 3.0, 3.5 and 4.0 magnitude levels. We divided the region containing L328 into two fields, F1 and F2, as shown in Fig. \ref{fig:328img}. The angular extend of L328 ($\sim1^{\prime}$) is found to be too small  to divide the field containing the cloud further into smaller sub-fields with sufficient number of stars for the analysis. Therefore we included the field containing an another molecular core, L323, located in the close vicinity ($\sim20^{\prime}$) of L328. Both the cores, L328 and L323, are found to have similar radial velocities of 6.6 and 6.4 km~s$^{-1}$ respectively \citep{1988ApJS...68..257C}. The locations of the cores and the fields chosen are identified and labelled in the Fig. \ref{fig:328img}. The stars, classified as dwarfs, that are used to estimate the core distance are identified using circles.

In Fig. \ref{fig:328alldist} we show the $A_{V}$ vs. $d$ plot for the stars from all  the fields (F1-2) combined together. The stars from F1 and F2 are represented using open and filled circles respectively. The field containing L328 is small when compared to F2 due to the small angular extend of the cloud. The stars showing high extinction at/or beyond 217 pc are present in both the fields. There exist two stars with  $A_{V}\geq1$ magnitude at distances smaller than 217 pc. Both stars are found to be located towards F2 at different distances. On the basis of 347 sources classified as main sequence stars from the two fields, we estimated a distance of $217\pm30$ pc to both L328 and L323.

Earlier estimates of distances to L323 (and L328) were by \citet{1974AJ.....79...42B} who assumed a distance of 200 pc based on the number of stars brighter than $m_{pg}=21$ mag projected on the cloud in their photographic plate. Later, several authors have assigned 200 pc assuming that the clouds, L328 and L323, are located in between the Ophiuchus cloud and the Aquila Rift \citep[e.g.,][]{2009ApJ...693.1290L}. Using photometry in \textit{Vilnius} system, \citet{2003A&A...405..585S} estimated a distance of $225\pm55$ pc to the front edge of the Aquila Rift with a possible depth of 80 pc to the cloud complex. If L328 and L323 are physically associated with the Aquila Rift, then possibly, the clouds are located to the front edge of it.

\subsubsection{L673-7}

\begin{figure}
\centering
\resizebox{9cm}{9cm}{\includegraphics{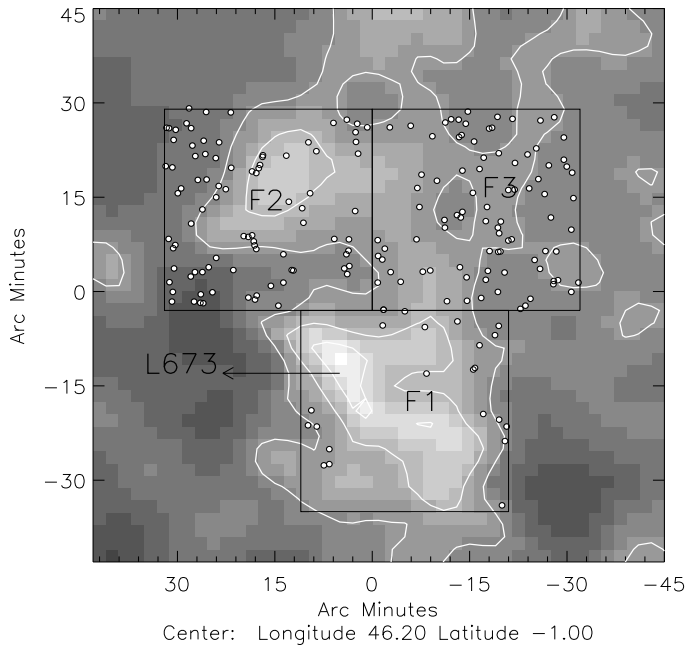}}
\caption{The $1.5^{\circ}\times1.5^{\circ}$ extinction map produced by \citet{2005PASJ...57S...1D} of the region containing L673-7. The contours are drawn at 4.0 and 5.0 magnitude levels. The fields used for selecting star to determine distance are identified and labelled. The stars classified as dwarfs using the near-IR photometry in each field are shown using filled circles. Each field covered an area is $ 0.5^{\circ}\times0.5^{\circ} $.\label{fig:673img}}
\end{figure}
\begin{figure}
\centering
\resizebox{9cm}{6.5cm}{\includegraphics{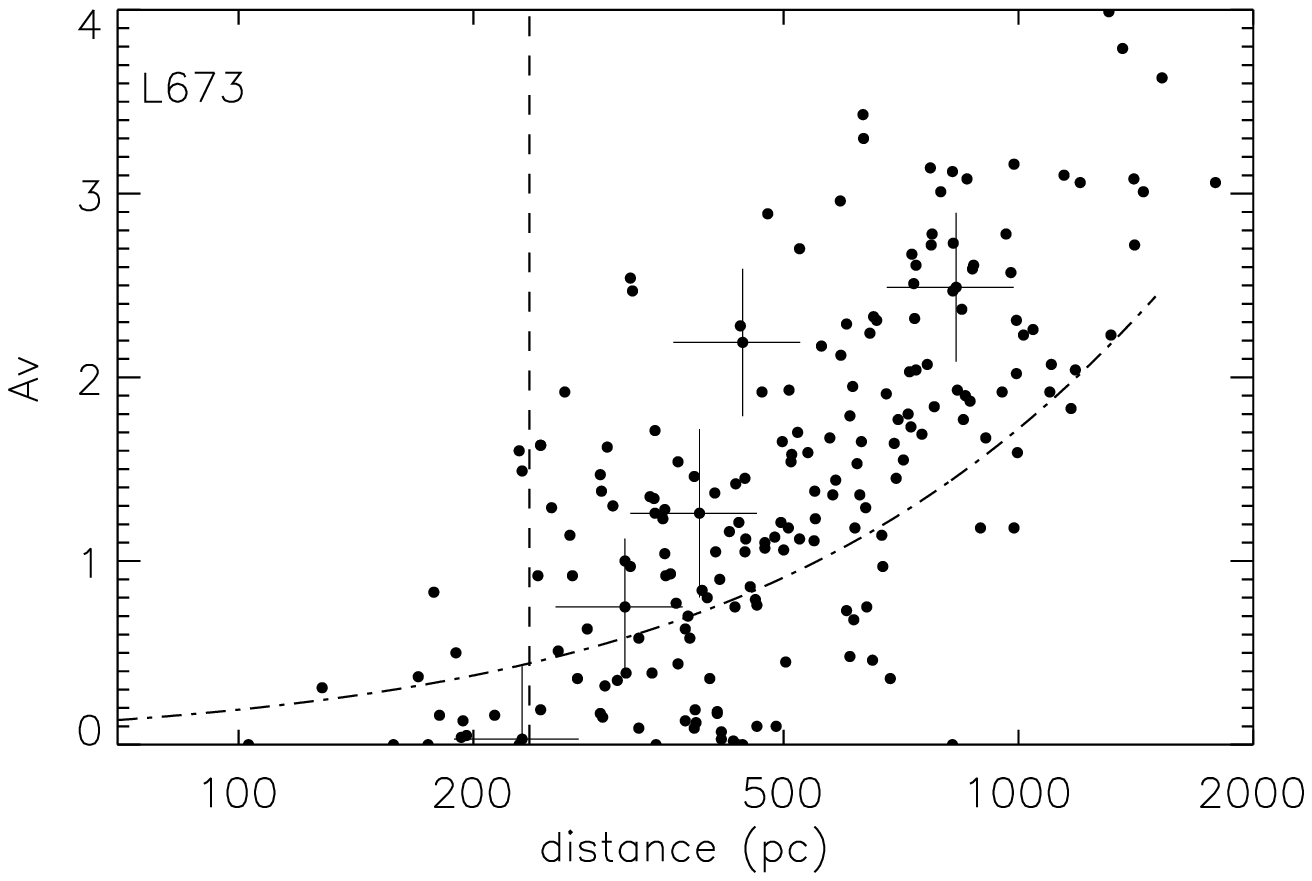}}
\caption{The $A_{V}$ vs. $d$ plot for all  the stars classified as dwarfs from the fields F1, F2 \& F3 combined together towards L673. The dashed vertical line is drawn at 240 pc inferred from the procedure described in the \S \ref{sec:data} (see Fig.\ref{fig:hist}). The dash-dotted curve represents  the increase in the extinction towards the Galactic latitude of $b=-1.33^{\circ}$ as a function of distance  produced from the expressions given by BS80. The error bars are shown on a few stars for better clarity of the points.\label{fig:all673dist}}
\end{figure}
\begin{figure}
\centering
\resizebox{9cm}{13cm}{\includegraphics{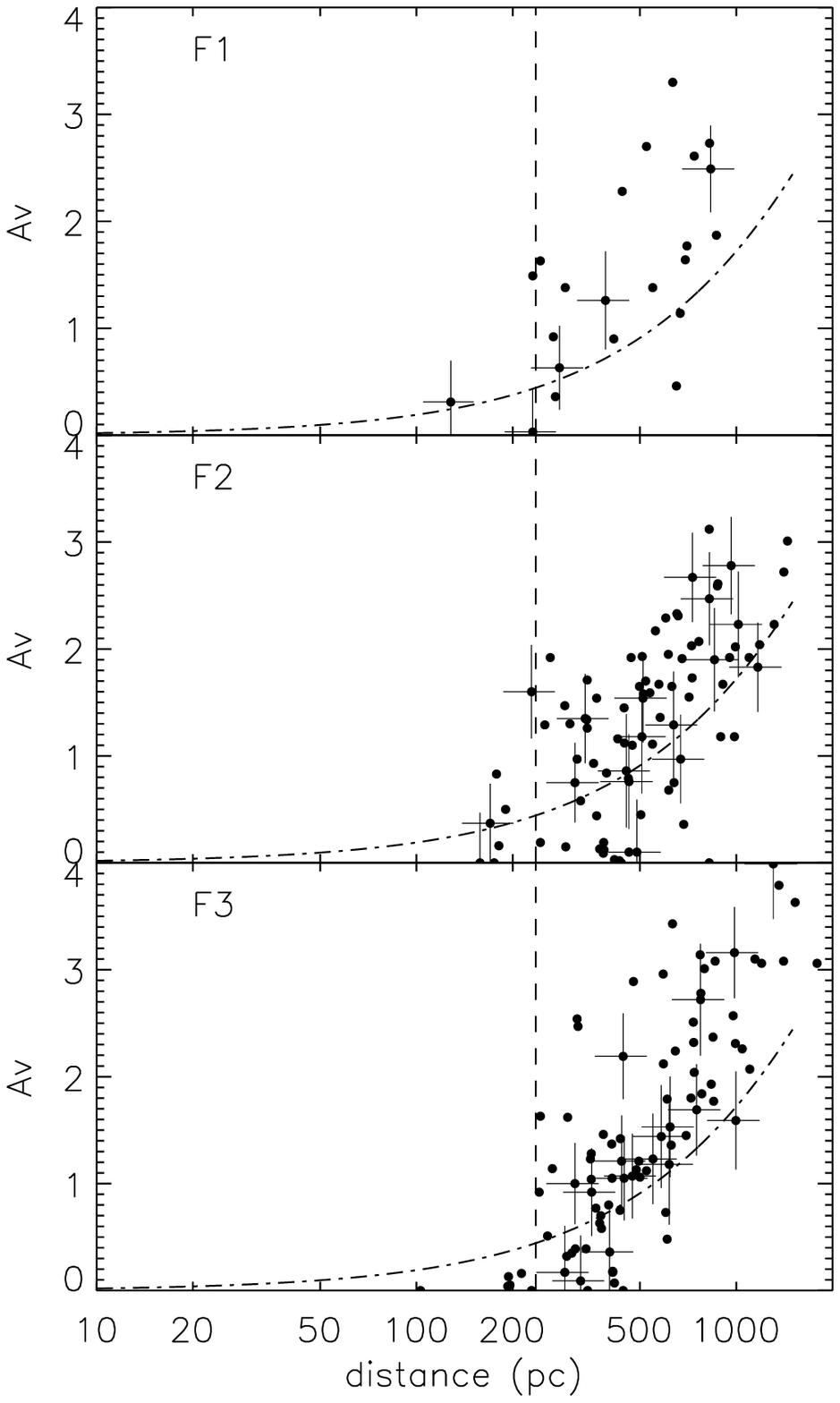}}
\caption{The $A_{V}$ vs. $d$ plot for the stars classified as dwarfs from the individual fields F1 (upper), F2 (middle) and F3 (lower panel) towards L673. The dashed vertical line is drawn at 240 pc inferred from the procedure described in the \S \ref{sec:data} (see Fig.\ref{fig:hist}). The dash-dotted curve represents  the increase in the extinction towards the Galactic latitude of $b=-1.33^{\circ}$ as a function of distance  produced from the expressions given by BS80. The error bars are shown on a few stars for better clarity of the points.\label{fig:673dist}}
\end{figure}

In Fig. \ref{fig:673img} we present the $1.5^{\circ}\times1.5^{\circ}$ extinction map towards the direction of L673-7. The contours are drawn at 4.0 and 5.0 magnitude levels. We divided the region containing L673-7 into three fields, F1, F2 and F3, as shown in Fig. \ref{fig:673img}.  L673-7 is part of the L673 cloud complex from the \citet{1962ApJS....7....1L} catalog. \citet{1999ApJS..123..233L}  identified 11 cores within this complex including L673-7. The radial velocity of L673 complex in CO line is shown to range from 6.71 - 7.30 km~s$^{-1}$ \citep{1983ApJ...265..766K}. Some more clouds are found to be associated with the region selected around L673 namely, L675, L676 from the \citet{1962ApJS....7....1L} catalog and CB188 from \citet{1988ApJS...68..257C}. These three clouds also show similar radial velocities of 7.5, 7.6 and 7.1 km~s$^{-1}$ respectively \citep{1988ApJS...68..257C} as that of L673 complex. We therefore assume that the regions selected around L673 are all associated and included in our analysis. The stars, classified as dwarfs, that are used for determining the cloud distance are identified using circles.

In Fig. \ref{fig:all673dist} we present the $A_{V}$ vs. $d$ plot for the stars from all  the fields (F1-3) combined together. A jump in the extinction values significantly above the values expected from the expression of BS80 is evident at/or beyond $\sim240$ pc. The stars showing high extinction are distributed almost uniformly beyond $\sim240$ pc. In Fig. \ref{fig:673dist}, we show the $A_{V}$ vs. $d$ plot for the stars from the individual fields F1, F2 and F3. In all  the three fields, the jump in the extinction is evidently occurring at $\sim240$ pc. Using 199 sources classified as main sequence stars from the three fields, we estimated a distance of $240\pm45$ pc to L673-7.

Previous determination of distances to L673-7 were highly uncertain. \citet{1983AJ.....88.1040H} used a distance of 300 pc, based on proper motion studies. Others assumed a distance of either 200 pc or 300 pc in their studies \citep[e.g.,][]{1995A&AS..113..325H, 2010ApJ...721..995D}. \citet{1982ApJ...261..151E} used a distance of 150 pc based on the assumption that L673 is a part of Gould's belt and characterized it as the minimum distance to the cloud. Because L673, L675, L676 and CB188 show similar radial velocities, we assign a distance of $240\pm45$ pc to all these clouds.
\subsubsection{L1148}
\begin{figure*}
\centering
\resizebox{15cm}{10cm}{\includegraphics{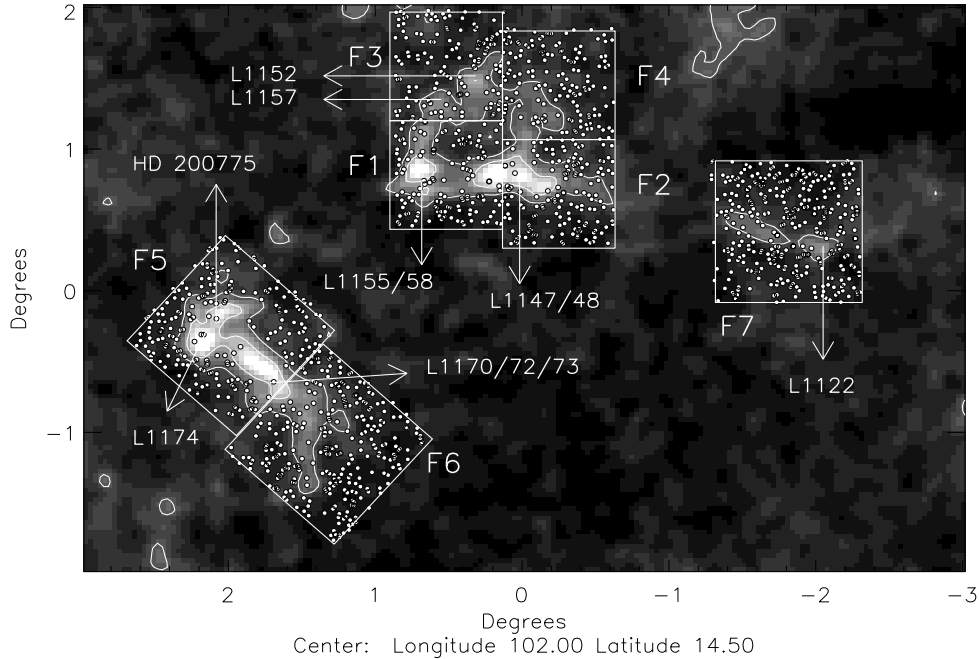}}
\caption{The $ 2^{\circ}\times2^{\circ} $ extinction map produced by \citep{2005PASJ...57S...1D} of the region containing LDN 1148. The contours are drawn at 0.7, 1.0 and 2.0 magnitude levels. The fields used for selecting star to determine distance are identified and labelled. Each field is of  $ 0.7^{\circ}\times0.7^{\circ} $.\label{fig:1148img}}
\end{figure*}
\begin{figure}[h]
\centering
\resizebox{9cm}{6.5cm}{\includegraphics{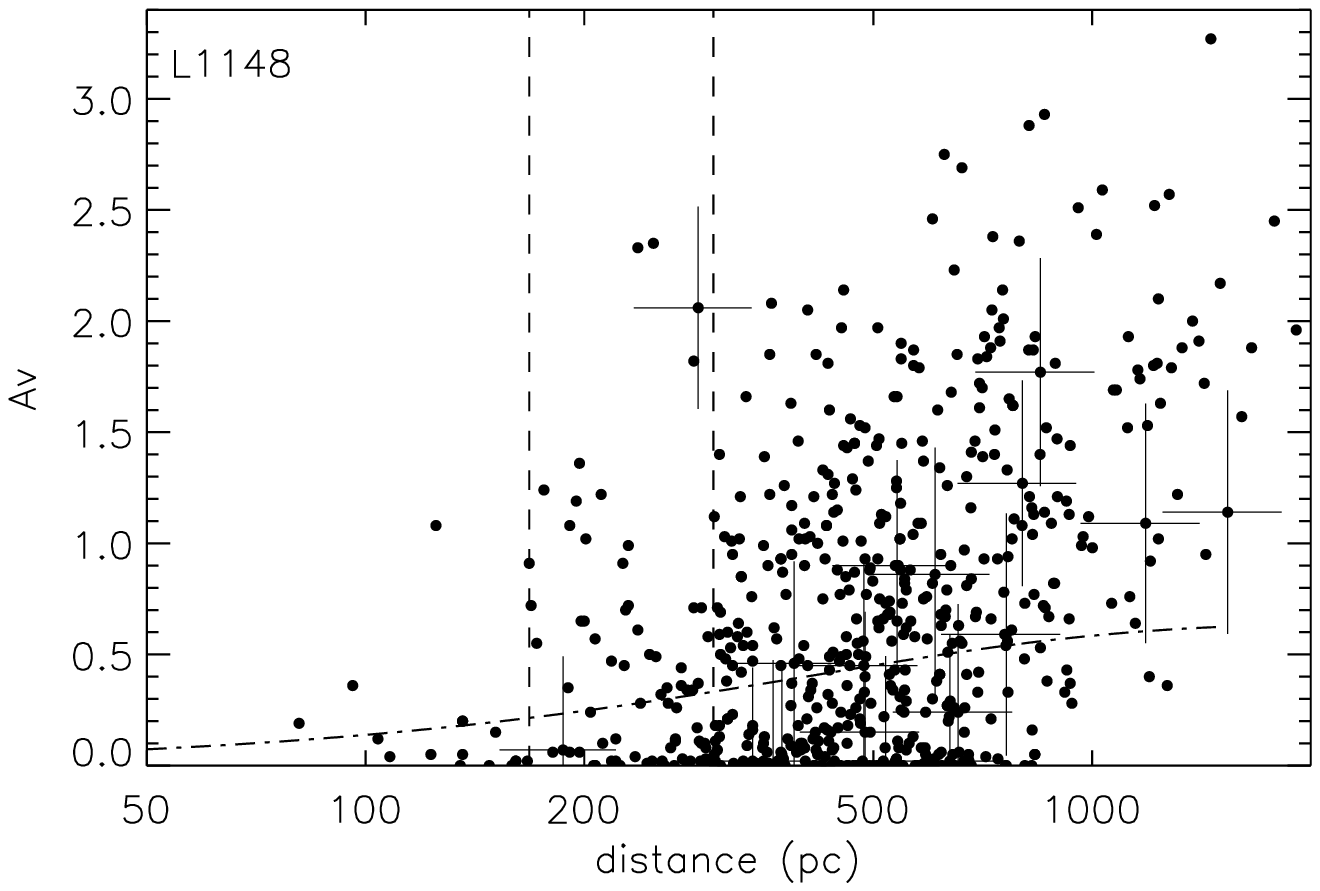}}
\caption{The $A_{V}$ vs. $d$ plot for all the stars classified as dwarfs from the fields F1, F2, F3, \& F4 combined together towards L1148. The dashed vertical line is drawn at 168 pc inferred from the procedure described in the \S \ref{sec:data} (see Fig.\ref{fig:hist}). The dash-dotted curve represents  the increase in the extinction towards the Galactic latitude of $b=+15.3^{\circ}$ as a function of distance  produced from the expressions given by BS80. The error bars are not shown on all the stars for better clarity. The sources shown using filled circles in red are those identified with black squares in Fig. \ref{fig:1148img}.\label{fig:all1148dist}}
\end{figure}
\begin{figure}
\centering
\resizebox{9cm}{12cm}{\includegraphics{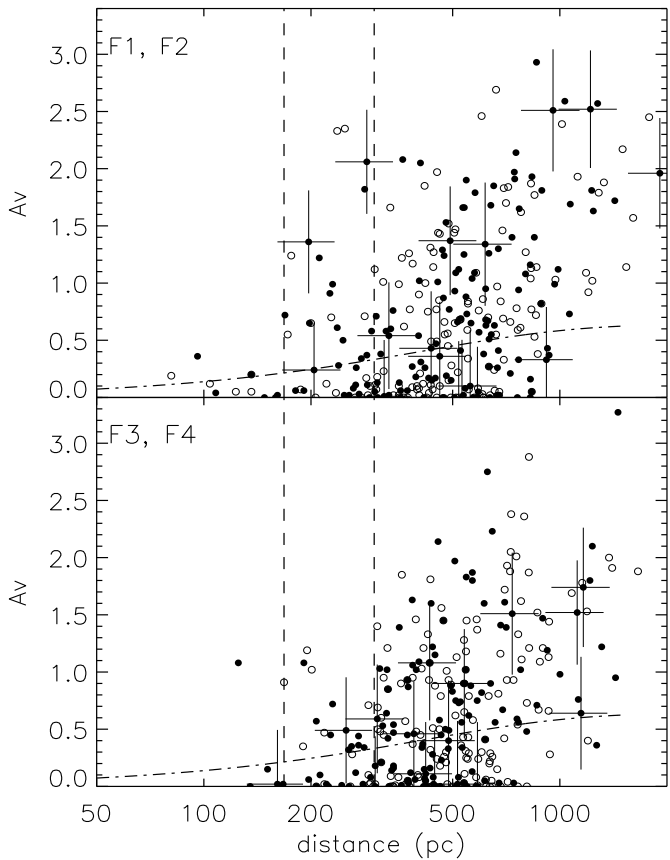}}
\caption{The $A_{V}$ vs. $d$ plot for the stars classified as dwarfs from the individual fields F1 (open circles) and F2 (filled circles) in upper panel and F3 (filled circles) and F4 (open circles) in lower panel towards L1148. The dashed vertical lines are drawn at 168 pc and 301 pc inferred from the procedure described in the \S \ref{sec:data} (see Fig.\ref{fig:hist}). The dash-dotted curve represents  the increase in the extinction towards the Galactic latitude of $b=+15.3^{\circ}$ as a function of distance  produced from the expressions given by BS80. The error bars are shown on a few stars for better clarity of the points.\label{fig:1148dist}}
\end{figure}
\begin{figure}
\centering
\resizebox{9cm}{12cm}{\includegraphics{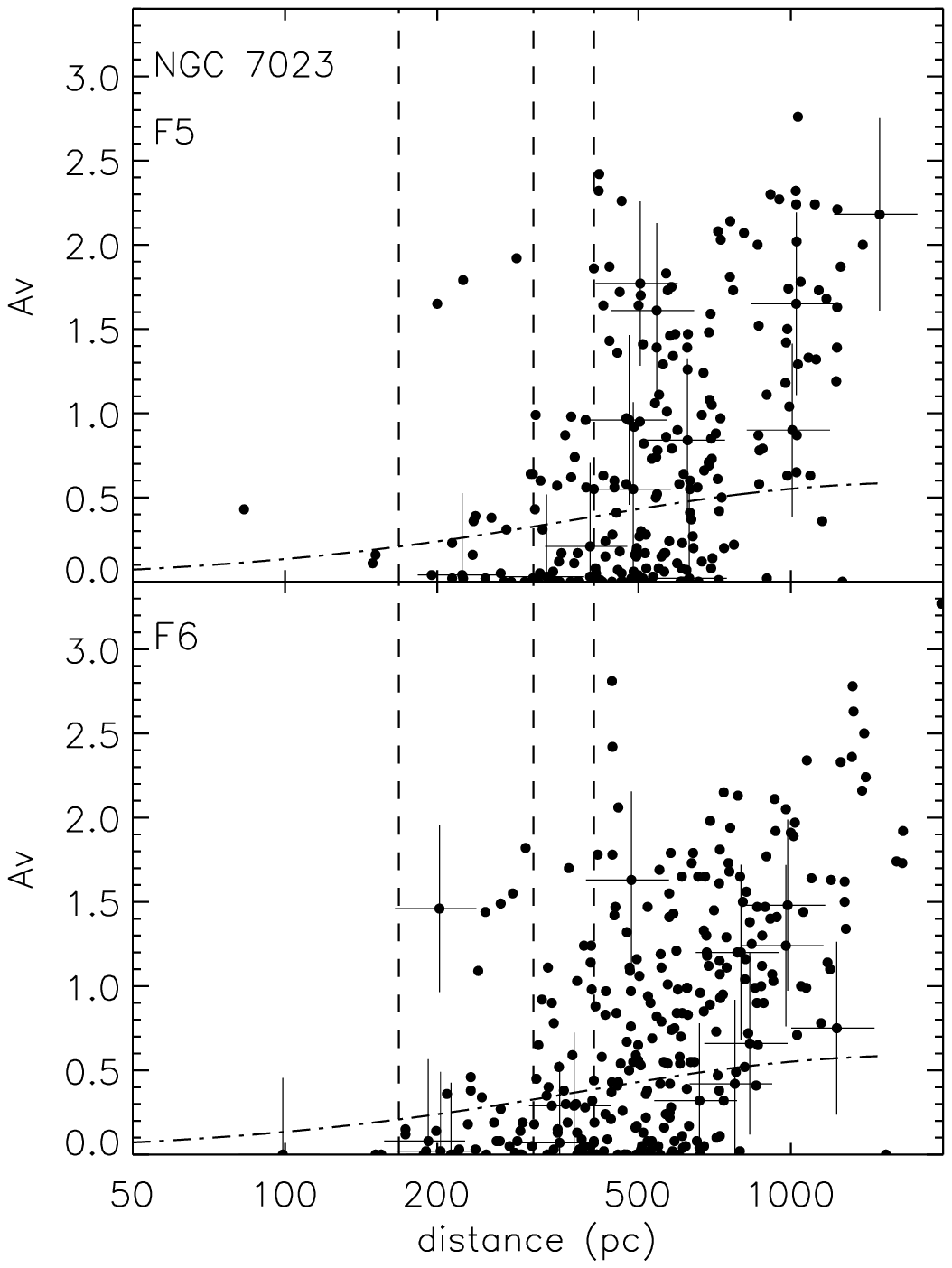}}
\caption{The $A_{V}$ vs. $d$ plot for the stars classified as dwarfs from the individual fields F5 (upper panel) and F6 (lower panel) towards NGC 7023. The dashed vertical lines are drawn at 168 pc, 310 pc and 408 pc inferred from the procedure described in the \S \ref{sec:data} (see Fig.\ref{fig:hist}). The dash-dotted curve represents  the increase in the extinction towards the Galactic latitude of $b=+14.1^{\circ}$ as a function of distance  produced from the expressions given by BS80. The error bars are shown on a few stars for better clarity of the points.\label{fig:7023dist}}
\end{figure}

The fields used to determine the distance to L1148 are shown on the extinction map in Fig. \ref{fig:1148img}. The contours are drawn at 0.7, 1.0 and 2.0 magnitude levels. We divided the region containing L1147/48 into four fields, F1, F2, F3 and F4, as identified on Fig. \ref{fig:1148img}. The stars, classified as dwarfs, that were used to determine the cloud distance are identified using circles.

In Fig. \ref{fig:all1148dist} we present the $A_{V}$ vs. $d$ plot for the stars from all the fields (F1-4) combined together. A jump in the extinction values significantly above the values expected from the expression of BS80 is evident at 168 pc. But if we compare the $A_{V}$ vs. $d$ plots of IRAM04191 and L1521F, which are both located within 200 pc distance, with that of L1148, we can notice that the stars with high extinction towards L1148 are not distributed continuously beyond 168 pc.  The angular size of the cloud (inferred from the extinction map shown in Fig. \ref{fig:1148img}) is large enough to have shown a continuous distribution of stars with high extinction beyond 168 pc if the cloud would have been located at this distance. The lack of such an effect is clearly seen in Fig. \ref{fig:1148dist} where we show the $A_{V}$ vs. $d$ plot for the stars from the individual fields F1 (open circles) and F2 (filled cicles) in the upper panel and F3 (filled circles) and F4 (open circles) in the lower panel. Only few stars are seen with high extinction especially towards the fields F3 and F4. On the contrary, a wall of high extinction stars are seen beyond $\sim300$ pc towards all the four fields. Using the stars from F3 and F4 alone we estimated a distance of $301\pm55$ pc to this dust component.

The L1148 core is located in a cloud complex towards the direction of Cepheus. The most widely quoted distance to L1147/48/55 is $325\pm13$ pc \citep{1992BaltA...1..149S}. They used the Vilnius photometric system to classify the stars and also to get the interstellar reddening. They obtained a distance of $325\pm13$ pc to L1147/1158 by taking an average of ten stars showing $A_{V}\geq0.45$ which were distributed in the range 240-380 pc. \citet{1981ApJS...45..121S} obtained a distance of 250 pc to L1147 based on the $A_{V}$ vs. $d$ plot produced using the stars from a region of $20^{\circ}\times20^{\circ}$ centred on the cloud that had both optical photometry and MK spectral classification. \citet{1998ApJS..115...59K} showed the presence of three dust components at characteristic distances: 200, 300, and 450 pc, based on cumulative distribution of field star distance moduli in Wolf diagrams. A complex velocity structure was reported by \citet{1991A&A...249..493H} towards the direction of L1147/48, L1152, L1157 and L1155/58 cloud complex. They found the presence of two distinct velocity components in $^{13}$CO and C$^{18}$O line profiles. The blueshifted component ($ V_{LSR}\approx1.5$~km~s$^{-1} $) was found to be limited to smaller extend whereas the redshifted component ($V_{LSR}\approx2-4$~km~s$^{-1} $) was found to be extended over the whole cloud area. Based on the velocity structure found over the whole cloud, they suggested the presence of two cloud components along the line of sight. 

The radial velocities measured for the cores L1148, L1155C-1 and L1155C-2 of the cloud complex are found to be 2.56, 2.58 and 1.35~km~s$^{-1} $ respectively \citep{2004ApJS..153..523L}. Another cloud, L1167/1174, which also showed similar radial velocity \citep[2.67~km~s$^{-1}$, ][]{2002ApJ...572..238C, 1997ApJS..110...21Y} as that of L1147/48 is found to be located  at $\sim2^{\circ}$ south-east of L1147/48/55 cloud complex (see Fig. \ref{fig:1148img}). \citet{1992BaltA...1..149S} estimated a distance of $288\pm25$ pc to the L1167/1174 again by taking an average distance of four considerably reddened stars. They preferred a distance of 275 pc to HD 200775, a Herbig Be star \citep{1994A&AS..104..315T} found associated with NGC 7023 and responsible for the reflection nebulosity, by assuming it to be a B3Ve star. \citet{2010A&A...509A..44M}, using the same method as employed in this work, have estimated a distance of $408\pm76$ pc to NGC 7023 \citep[consistent with the Hipparcos parallax distance of $429^{+156}_{-90}$ pc to HD 200775,][]{1998A&A...330..145V, 1999A&A...352..574B}. In the direction of NGC 7023, they also found evidence for two additional dust components one at $\sim200$ pc and another at 305 pc which were in good agreement with those inferred by \citet{1998ApJS..115...59K}. In Fig. \ref{fig:7023dist} we show the $A_{V}$ vs. $d$ plot for the stars from the fields F5 and F6 which contain L1167/1174 complex (see Fig. \ref{fig:1148img}). Step-like features at 310 and 408 pc are clearly visible in the plots. The component at 408 pc seems to be more conspicuous towards F5 where L1174 cloud is located. A continuous distribution of stars with high extinction is apparent beyond 408 pc. HD 200775 is probably associated with this cloud component. Towards F6, the dust component found at 310 pc seems to be dominant. There are a number of stars both towards F5 and F6 that show high extinction even at distances smaller that 310 pc. Presence of a third dust layer in the line of sight could be the reason.

\begin{figure}
\centering
\resizebox{9cm}{6.5cm}{\includegraphics{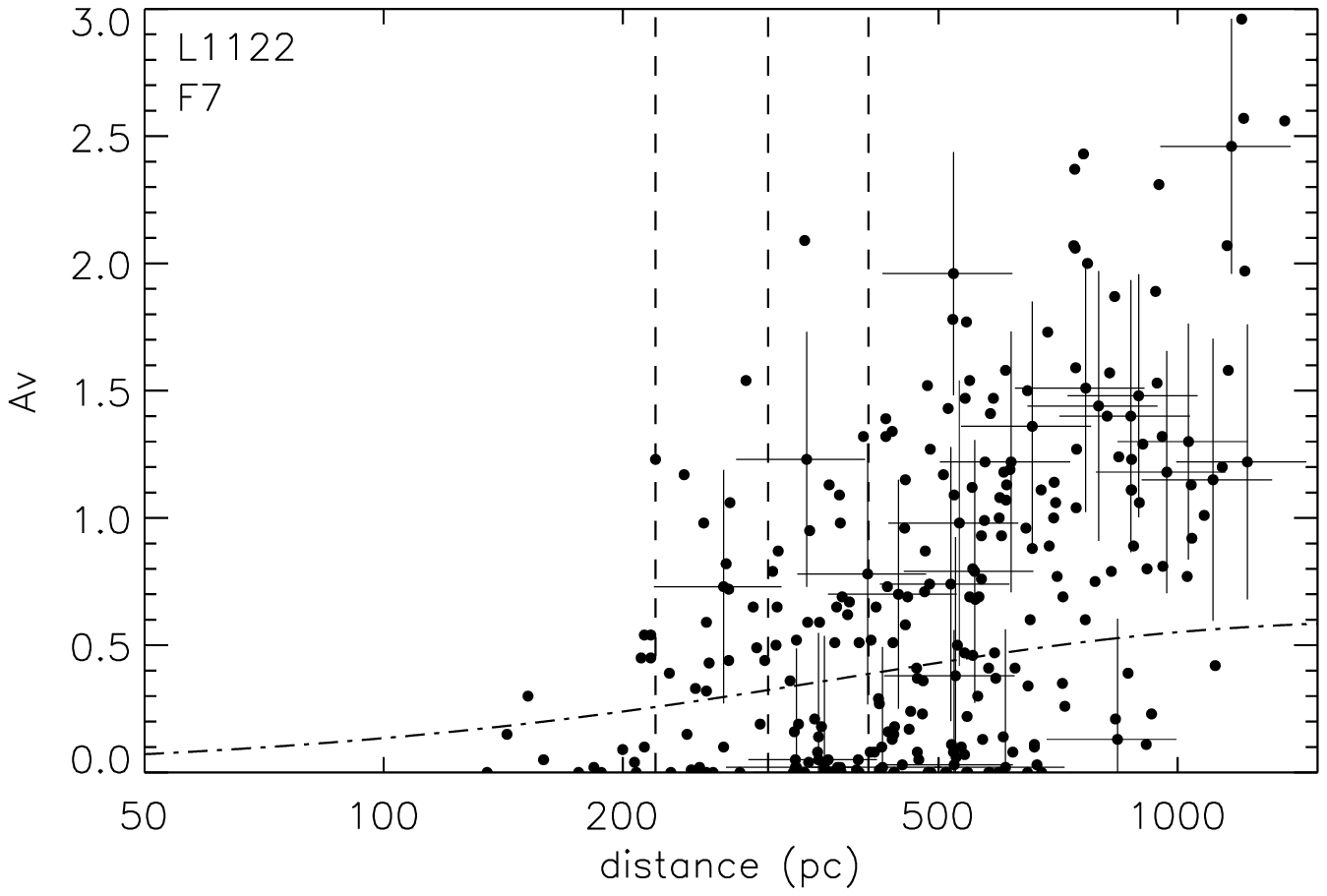}}
\caption{The $A_{V}$ vs. $d$ plot for all the stars classified as dwarfs towards L1122. The dashed vertical lines are drawn at 220, 301 and 408 pc. The dash-dotted curve represents  the increase in the extinction towards the Galactic latitude of $b=+14.8^{\circ}$ as a function of distance  produced from the expressions given by BS80. The error bars are shown on a few stars for better clarity of the points.\label{fig:1122dist}}
\end{figure}

About $2^{\circ}$ south-west of the L1147/48 cloud complex, there exist one more Lynds cloud, L1122. This cloud showed a radial velocity of $4.8$~km~s$^{-1}$ \citep{1997ApJS..110...21Y}. No earlier distance estimates are available for the cloud. We estimated distance to this cloud by selecting stars from a region of $1^{\circ}\times1^{\circ}$ (see Fig. \ref{fig:1148img}) centred on the cloud. In Fig. \ref{fig:1122dist}, we show the $A_{V}$ vs. $d$ plot for the stars towards L1122. A sudden jump in the extinction is visible at 220 pc. We assigned the distance of the first star that showed $A_{V}>1$ as the distance to this cloud. It should be noticed that the dust components which were seen at 310 and 408 pc towards L1167/1174 are absent towards L1122. The location of this cloud at a smaller distance proves that there exist dust components that are as close as 200 or even less as we found towards L1148. We therefore believe that the dust component seen towards L1147/48 cloud complex is a foreground layer and that the actual layer is associated with the dust component inferred at 301 pc.

\subsubsection{L1014}

\begin{figure}
\centering
\resizebox{9cm}{9cm}{\includegraphics{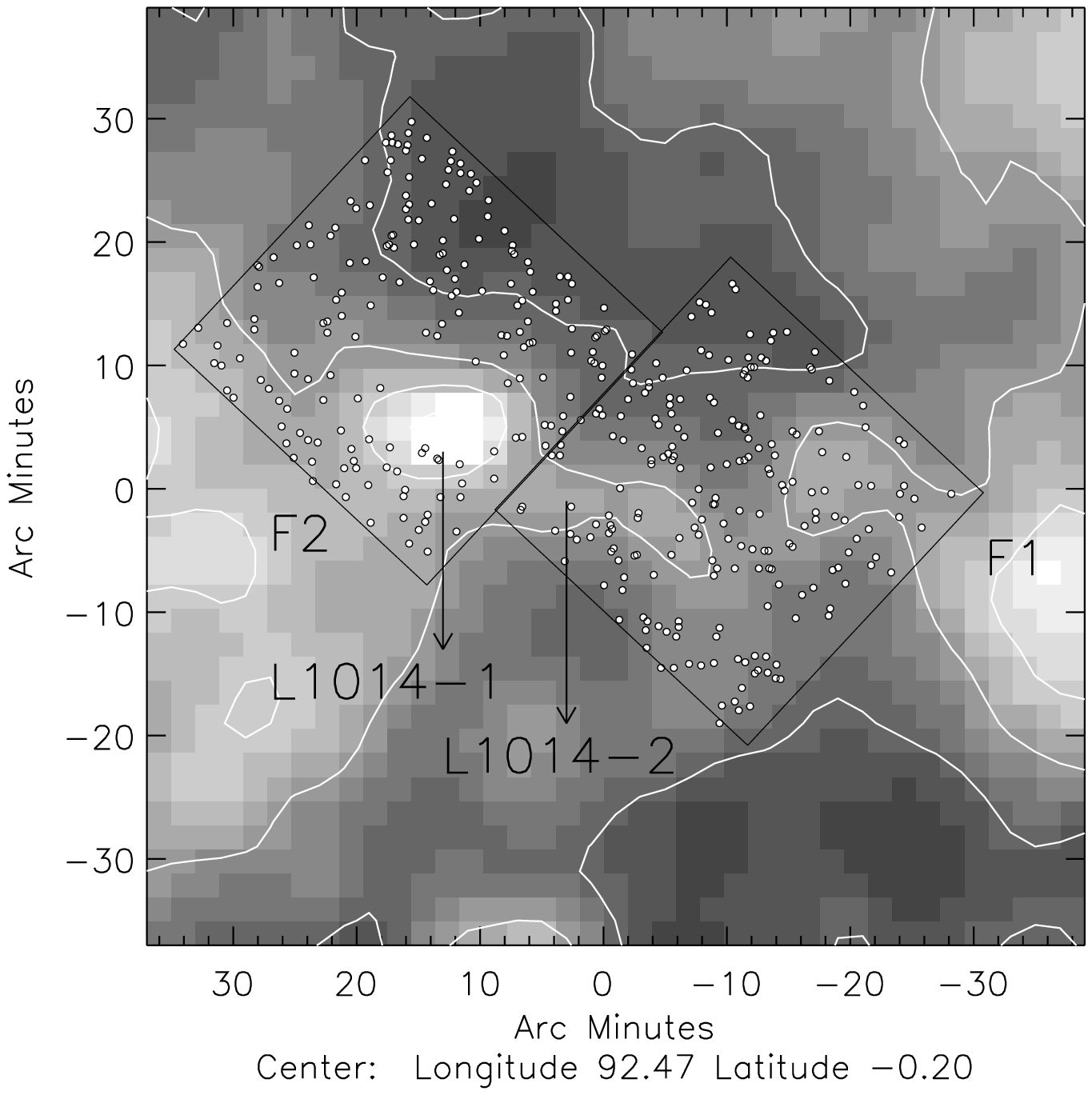}}
\caption{The $1.2^{\circ}\times1.2^{\circ}$ extinction map produced by \citet{2005PASJ...57S...1D} of the region containing L1014. The contours are drawn at 1.0, 1.5 and 2.0 magnitude levels. The fields used for selecting star to determine distance are identified and labelled. The stars classified as dwarfs using the near-IR photometry in each field are shown using filled circles. Each field is of  $ 0.4^{\circ}\times0.4^{\circ} $.\label{fig:1014img}}
\end{figure}
\begin{figure}

\resizebox{9cm}{6.5cm}{\includegraphics{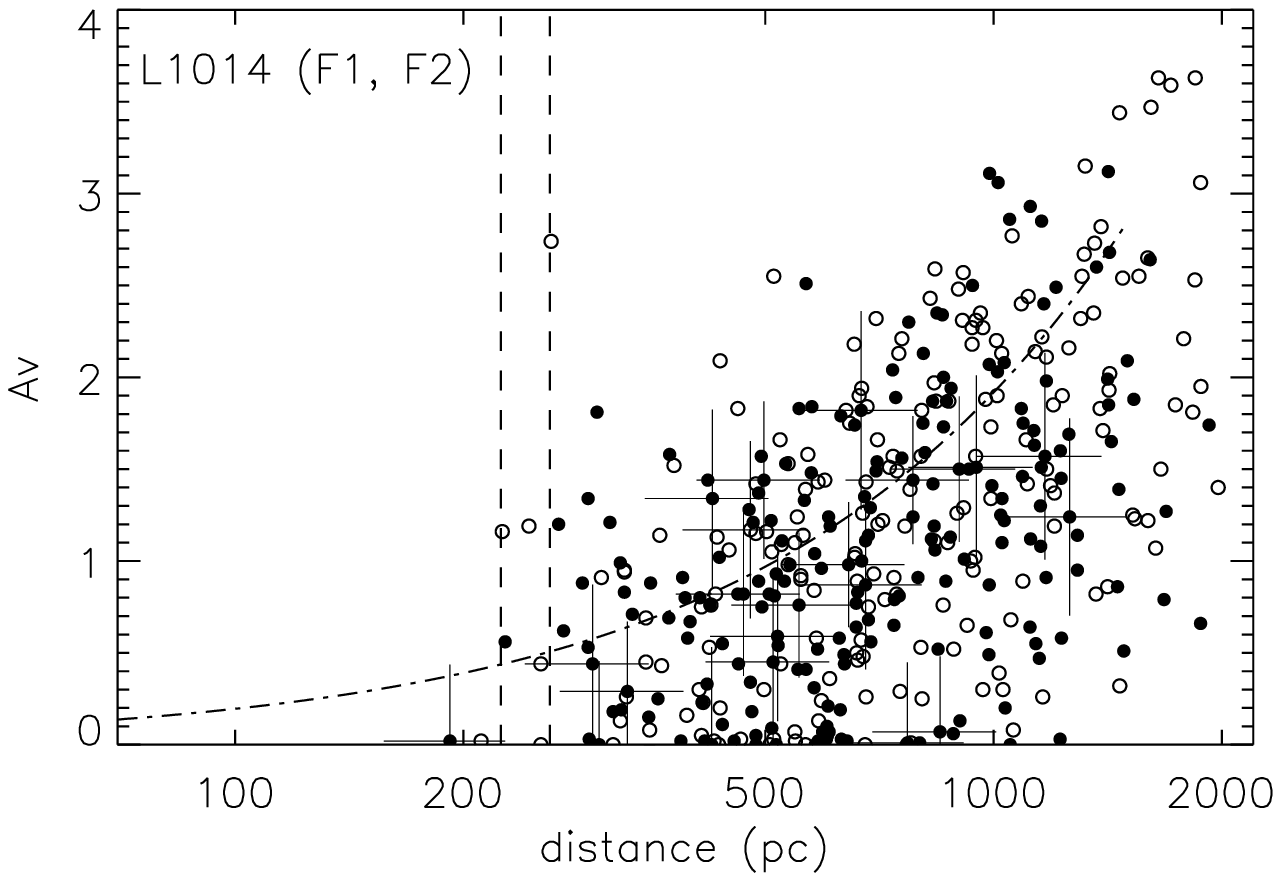}}
\caption{The $A_{V}$ vs. $d$ plot for all  the stars classified as dwarfs from the fields F1 \& F2 combined together towards L1014. The dashed vertical lines are drawn at 221 and 258 pc inferred from the procedure described in the \S \ref{sec:data} (see Fig.\ref{fig:hist}). The dash-dotted curve represents  the increase in the extinction towards the Galactic latitude of $b=-0.251^{\circ}$ as a function of distance  produced from the expressions given by BS80. The error bars are shown on a few stars for better clarity of the points.\label{fig:all1014dist}}
\end{figure}

In Fig. \ref{fig:1014img} we identify the fields used for selecting stars on the $1.2^{\circ}\times1.2^{\circ}$ extinction map of L1014. The contours are drawn at 1.0, 1.5 and 2.0 magnitude levels. The stars, classified as dwarfs, that are used for determining the cloud distance are identified using circles. Each field covers an area of  $ 0.4^{\circ}\times0.4^{\circ} $. The angular extend of L1014-2, which contains the VeLLO, L1014$-$IRS, is only $\approx2^{\prime}$. Therefore we included another cloud L1014-1 which is located $\sim10^{\prime}$ north-east of L1014-2.  The cloud L1014-1 is better known as B362. Both L1014-1 and L1014-2 are found to have similar radial velocities \citep[4.0 and 4.2$~km~s^{-1}$ respectively, ][]{1999ApJ...526..788L, 2005A&A...439.1023C}.

In Fig. \ref{fig:all1014dist} we present the $A_{V}$ vs. $d$ plot for the stars from all  the fields F1 (filled circles) and F2 (open circles) combined together. In Fig. \ref{fig:hist}, the steep rise in the mean values of extinction are found to happen at distances 221 and 258 pc. But from Fig. \ref{fig:all1014dist} it is evident that the steep rise in the extinction at 221 pc is occurring mainly due to two stars that are located in the field, F2 alone. On the contrary, the steep rise in the mean value of extinction at 258 pc (see Fig. \ref{fig:all1014dist}) is caused due to the stars from both fields, F1 and F2 (filled and open circles). Therefore we associate the jump in the extinction at 258 pc to the presence of the cores L1014-1/2. Also, beyond this distance the high extinction stars are distributed almost regularly. However, we found only two stars with low extinction values lying foreground to L1014-1/2 that are essential to have a complete census of the dust layers in the direction. In order to include more cores in the vicinity of L1014-1/2 that share similar radial velocities, we looked at the large-scale maps of the clouds produced by \citet{1994ApJS...95..419D} based on $^{13}$CO (with a $2.7^{\prime}$ angular resolution) survey towards the Cygnus region. L1014-1, the cloud which was detected in their survey, is found to be isolated with no cloud components, at least, within a radius of $2^{\circ}$ in the velocity range of $3<V_{LSR}<4.5~km~s^{-1}$ (see Fig. $5i$ of their publication). 

 \cite{1985ApJ...297..751D} have assigned a distance of 800 pc to L1014 assuming it to be part of Cyg OB7 \citep[see][]{1994ApJS...95..419D}. \cite{1978ApJ...221L.117H} adopted a distance of 200 pc for B362. \citet{2006PASJ...58L..41M} assigned a distance of 400-900 pc to L1014 by assuming possible association of identified T Tauri stars with the cloud. No reliable estimate of distance to L1014-2 is available in the literature. In most of the studies carried out on L1014-2, the authors have adopted a distance of 200 pc to it \citep[e.g.,][]{1999ApJS..123..233L} by assuming that both B362 and L1014-2 are located at same distance. Using a total of 400 sources classified as main sequence stars from the two fields, we estimated a distance of $258\pm50$ pc to L1014.


\subsection{Are VeLLOs really VeLLOs?\label{sec:vello}}
\begin{table}
\begin{center}
\caption{On the status of the VeLLOs.}\label{tab:vello}
\begin{tabular}{lcllcc}\hline
Cores							&$d^{\dagger}$&$L_{int}$		&Ref.					&$d^{\ddagger}$ 		&	$L_{int}$		\\
 			       					&(pc)              	&($L_{\odot}$)	&							&(pc)							&	$(L_{\odot}$)	\\\hline
IRAM04191						&140				&0.08				&1						&$127\pm25$			&$0.07	$\\ 
L1521F       					&140				&0.05				&2						&$136\pm36$			&$0.05  $\\  
BHR 111 						&250   				&0.04				&3						&$355\pm65$			&$0.08  $\\    					
L328			 				&200              	&0.06				&4						&$217\pm30$			&$0.07  $\\
L673-7         					&300				&0.04				&5						&$240\pm45$			&$0.03 $\\
L1014         					&200          		&0.09	  			&6						&$258\pm50$			&$0.15  $\\
L1148							&325				&0.10				&7						&$301\pm55$			&$0.09  $\\
\hline
\end{tabular}
\end{center}
$^{\dagger}$Distances used by authors to evaluate internal luminosities of the VeLLOs prior to this work.
$^{\ddagger}$Distance estimates from this work.\\
1. \citet{2006ApJ...651..945D}; 2. \citet{2006ApJ...649L..37B}; 3. \citet{2008ApJS..179..249D}; 4. \citet{2009ApJ...693.1290L}; 5. \citet{2010ApJ...721..995D}; 6. \citet{2004ApJS..154..396Y}; 7. \citet{2005AN....326..878K}; 
\end{table}
Using the distances estimated to the seven cores in this work we re-evaluate the internal luminosities of the VeLLO candidates. In Table \ref{tab:vello} we list the distances used in earlier studies and the internal luminosities calculated using those distances of the VeLLOs in columns 2 and 3 respectively. The re-evaluated internal luminosities of the VeLLOs are given in column 6 of the Table \ref{tab:vello}. The previously determined internal luminosities of the VeLLOs in the cores L328, L673-7 and L1014 were calculated using  distances that were highly uncertain.  However, we find that these distances are very close to the values we have determined in this work. As we can see from the Table \ref{tab:vello} that the re-estimation of internal luminosities of six VeLLO candidates with our reliable distance measurements confirm them to be consistent with the definition of the VeLLOs. The VeLLO associated with L1014-2 core is found to be $L_{int}=0.15$ $L_{\odot}$. However, this source is still interesting and could be called as VeLLO-like since the luminosity is still an order of magnitude less than the accretion luminosity of $L_{acc}\sim1.7$ $L_{\odot}$  estimated for a 0.08 $M_{\odot}$ protostar using an accretion rate of $\sim2\times10^{-6}$ $M_{\odot}$ yr$^{-1}$ \citep[the rate predicted by the standard model;][]{1987ARA&A..25...23S} and 3 $R_{\odot}$ stellar radius in the expression, $L_{acc}=3.13\times10^{7}~M\dot{M}_{acc}/R$ $L_{\odot}$.

\section{Conclusions}\label{sec:conclude}
We estimated distances to seven dark clouds IRAM04191, L1521F, BHR111, L328, L673, L1014, and L1148 which are found to harbour VeLLO candidates, discovered in the Spitzer Legacy program $ - $ \textit{From Molecular Cores
to Planet Forming Disks}, using near-IR photometric method. In this method, first the 2MASS $JHK_{s}$ photometry are used to produce the ($J-H$) and ($H-K_{s}$) colours of the stars projected on the field containing each cloud. These observed colours are then dereddened simultaneously using trial values of $A_{V}$  and a normal interstellar extinction law.  The best fit of the dereddened colours to the intrinsic  colours giving a minimum value of $\chi^{2}$ then yields the corresponding spectral type and $A_{V}$ for the star. The main sequence stars, thus classified, are then utilized in an $A_{V}$ versus distance plot to bracket the cloud distances. The typical error in the estimation of distances to the clouds are found to be $\sim18\%$.  Our distance estimates are: $127\pm25$ pc (IRAM04191), $136\pm36$ pc (L1521F), $355\pm65$ pc (BHR111), $ 217\pm30 $ pc (L328), $ 240\pm45 $ pc (L673), $ 258\pm50 $ pc (L1014), and $ 301\pm55 $ pc (L1148). Using these distance estimates, we re-evaluated the internal luminosities of the VeLLO candidates discovered in these cores. Except L1014$-$IRS ($L_{int}=0.15$ $L_{\odot}$), all other VeLLO candidates are found to be consistent with the definition of a VeLLO.

\section*{Acknowledgments}
We express our gratitude to the anonymous referee for insightful and thoughtful comments and suggestions that have helped us to improved the paper substantially. This research was supported by Basic Science Research Program through the National Research Foundation of Korea (NRF) funded  by the Ministry of Education, Science and Technology (2010-0011605). This publication makes use of data products from the Two Micron all Sky Survey, which is a joint project of the University of Massachusetts and the Infrared Processing and Analysis Center/California Institute of Technology, funded by the National Aeronautics and Space Administration and the National Science Foundation. This research has also made use of the SIMBAD database, operated at CDS, Strasbourg, France. 
\bibliographystyle{aa}
\bibliography{myref}

\begin{thebibliography}{83}
\expandafter\ifx\csname natexlab\endcsname\relax\def\natexlab#1{#1}\fi

\bibitem[{{Alves} \& {Franco}(2007)}]{2007A&A...470..597A}
{Alves}, F.~O. \& {Franco}, G.~A.~P. 2007, \aap, 470, 597

\bibitem[{{Bahcall} \& {Soneira}(1980)}]{1980ApJS...44...73B}
{Bahcall}, J.~N. \& {Soneira}, R.~M. 1980, \apjs, 44, 73

\bibitem[{{Bertout} {et~al.}(1999){Bertout}, {Robichon}, \&
  {Arenou}}]{1999A&A...352..574B}
{Bertout}, C., {Robichon}, N., \& {Arenou}, F. 1999, \aap, 352, 574

\bibitem[{{Bok} \& {McCarthy}(1974)}]{1974AJ.....79...42B}
{Bok}, B.~J. \& {McCarthy}, C.~C. 1974, \aj, 79, 42

\bibitem[{{Bourke} {et~al.}(2005){Bourke}, {Crapsi}, {Myers}, {Evans},
  {Wilner}, {Huard}, {J{\o}rgensen}, \& {Young}}]{2005ApJ...633L.129B}
{Bourke}, T.~L., {Crapsi}, A., {Myers}, P.~C., {et~al.} 2005, \apjl, 633, L129

\bibitem[{{Bourke} {et~al.}(1995{\natexlab{a}}){Bourke}, {Hyland}, \&
  {Robinson}}]{1995MNRAS.276.1052B}
{Bourke}, T.~L., {Hyland}, A.~R., \& {Robinson}, G. 1995{\natexlab{a}}, \mnras,
  276, 1052

\bibitem[{{Bourke} {et~al.}(1995{\natexlab{b}}){Bourke}, {Hyland}, {Robinson},
  {James}, \& {Wright}}]{1995MNRAS.276.1067B}
{Bourke}, T.~L., {Hyland}, A.~R., {Robinson}, G., {James}, S.~D., \& {Wright},
  C.~M. 1995{\natexlab{b}}, \mnras, 276, 1067

\bibitem[{{Bourke} {et~al.}(2006){Bourke}, {Myers}, {Evans}, {Dunham},
  {Kauffmann}, {Shirley}, {Crapsi}, {Young}, {Huard}, {Brooke}, {Chapman},
  {Cieza}, {Lee}, {Teuben}, \& {Wahhaj}}]{2006ApJ...649L..37B}
{Bourke}, T.~L., {Myers}, P.~C., {Evans}, II, N.~J., {et~al.} 2006, \apjl, 649,
  L37

\bibitem[{{Caselli} {et~al.}(2002){Caselli}, {Benson}, {Myers}, \&
  {Tafalla}}]{2002ApJ...572..238C}
{Caselli}, P., {Benson}, P.~J., {Myers}, P.~C., \& {Tafalla}, M. 2002, \apj,
  572, 238

\bibitem[{{Cernis}(1990)}]{1990Ap&SS.166..315C}
{Cernis}, K. 1990, \apss, 166, 315

\bibitem[{{Cernis} \& {Strai{\v z}ys}(1992)}]{1992BaltA...1..163C}
{Cernis}, K. \& {Strai{\v z}ys}, V. 1992, Baltic Astronomy, 1, 163

\bibitem[{{Clemens} \& {Barvainis}(1988)}]{1988ApJS...68..257C}
{Clemens}, D.~P. \& {Barvainis}, R. 1988, \apjs, 68, 257

\bibitem[{{Corradi} {et~al.}(1997){Corradi}, {Franco}, \&
  {Knude}}]{1997A&A...326.1215C}
{Corradi}, W.~J.~B., {Franco}, G.~A.~P., \& {Knude}, J. 1997, \aap, 326, 1215

\bibitem[{{Crapsi} {et~al.}(2005){Crapsi}, {Devries}, {Huard}, {Lee}, {Myers},
  {Ridge}, {Bourke}, {Evans}, {J{\o}rgensen}, {Kauffmann}, {Lee}, {Shirley}, \&
  {Young}}]{2005A&A...439.1023C}
{Crapsi}, A., {Devries}, C.~H., {Huard}, T.~L., {et~al.} 2005, \aap, 439, 1023

\bibitem[{{Cutri} {et~al.}(2003){Cutri}, {Skrutskie}, {van Dyk}, {Beichman},
  {Carpenter}, {Chester}, {Cambresy}, {Evans}, {Fowler}, {Gizis}, {Howard},
  {Huchra}, {Jarrett}, {Kopan}, {Kirkpatrick}, {Light}, {Marsh}, {McCallon},
  {Schneider}, {Stiening}, {Sykes}, {Weinberg}, {Wheaton}, {Wheelock}, \&
  {Zacarias}}]{2003tmc..book.....C}
{Cutri}, R.~M., {Skrutskie}, M.~F., {van Dyk}, S., {et~al.} 2003, {2MASS All
  Sky Catalog of point sources.}, ed. {Cutri, R.~M., Skrutskie, M.~F., van Dyk,
  S., Beichman, C.~A., Carpenter, J.~M., Chester, T., Cambresy, L., Evans, T.,
  Fowler, J., Gizis, J., Howard, E., Huchra, J., Jarrett, T., Kopan, E.~L.,
  Kirkpatrick, J.~D., Light, R.~M., Marsh, K.~A., McCallon, H., Schneider, S.,
  Stiening, R., Sykes, M., Weinberg, M., Wheaton, W.~A., Wheelock, S., \&
  Zacarias, N.}

\bibitem[{{Dame} \& {Thaddeus}(1985)}]{1985ApJ...297..751D}
{Dame}, T.~M. \& {Thaddeus}, P. 1985, \apj, 297, 751

\bibitem[{{Dobashi} {et~al.}(1994){Dobashi}, {Bernard}, {Yonekura}, \&
  {Fukui}}]{1994ApJS...95..419D}
{Dobashi}, K., {Bernard}, J., {Yonekura}, Y., \& {Fukui}, Y. 1994, \apjs, 95,
  419

\bibitem[{{Dobashi} {et~al.}(2005){Dobashi}, {Uehara}, {Kandori}, {Sakurai},
  {Kaiden}, {Umemoto}, \& {Sato}}]{2005PASJ...57S...1D}
{Dobashi}, K., {Uehara}, H., {Kandori}, R., {et~al.} 2005, \pasj, 57, 1

\bibitem[{{Dunham} {et~al.}(2008){Dunham}, {Crapsi}, {Evans}, {Bourke},
  {Huard}, {Myers}, \& {Kauffmann}}]{2008ApJS..179..249D}
{Dunham}, M.~M., {Crapsi}, A., {Evans}, II, N.~J., {et~al.} 2008, \apjs, 179,
  249

\bibitem[{{Dunham} {et~al.}(2010{\natexlab{a}}){Dunham}, {Evans}, {Bourke},
  {Myers}, {Huard}, \& {Stutz}}]{2010ApJ...721..995D}
{Dunham}, M.~M., {Evans}, N.~J., {Bourke}, T.~L., {et~al.} 2010{\natexlab{a}},
  \apj, 721, 995

\bibitem[{{Dunham} {et~al.}(2010{\natexlab{b}}){Dunham}, {Evans}, {Terebey},
  {Dullemond}, \& {Young}}]{2010ApJ...710..470D}
{Dunham}, M.~M., {Evans}, N.~J., {Terebey}, S., {Dullemond}, C.~P., \& {Young},
  C.~H. 2010{\natexlab{b}}, \apj, 710, 470

\bibitem[{{Dunham} {et~al.}(2006){Dunham}, {Evans}, {Bourke}, {Dullemond},
  {Young}, {Brooke}, {Chapman}, {Myers}, {Porras}, {Spiesman}, {Teuben}, \&
  {Wahhaj}}]{2006ApJ...651..945D}
{Dunham}, M.~M., {Evans}, II, N.~J., {Bourke}, T.~L., {et~al.} 2006, \apj, 651,
  945

\bibitem[{{Dzib} {et~al.}(2010){Dzib}, {Loinard}, {Mioduszewski}, {Boden},
  {Rodr{\'{\i}}guez}, \& {Torres}}]{2010ApJ...718..610D}
{Dzib}, S., {Loinard}, L., {Mioduszewski}, A.~J., {et~al.} 2010, \apj, 718, 610

\bibitem[{{Edwards} \& {Snell}(1982)}]{1982ApJ...261..151E}
{Edwards}, S. \& {Snell}, R.~L. 1982, \apj, 261, 151

\bibitem[{{Evans} {et~al.}(2003){Evans}, {Allen}, {Blake}, {Boogert}, {Bourke},
  {Harvey}, {Kessler}, {Koerner}, {Lee}, {Mundy}, {Myers}, {Padgett},
  {Pontoppidan}, {Sargent}, {Stapelfeldt}, {van Dishoeck}, {Young}, \&
  {Young}}]{2003PASP..115..965E}
{Evans}, II, N.~J., {Allen}, L.~E., {Blake}, G.~A., {et~al.} 2003, \pasp, 115,
  965

\bibitem[{{Franco}(2002)}]{2002MNRAS.331..474F}
{Franco}, G.~A.~P. 2002, \mnras, 331, 474

\bibitem[{{Gottlieb} \& {Upson}(1969)}]{1969ApJ...157..611G}
{Gottlieb}, D.~M. \& {Upson}, II, W.~L. 1969, \apj, 157, 611

\bibitem[{{Greenstein} \& {Shapley}(1937)}]{1937AnHar.105..359G}
{Greenstein}, J.~L. \& {Shapley}, H. 1937, Annals of Harvard College
  Observatory, 105, 359

\bibitem[{{Harjunp{\"a}{\"a}} {et~al.}(1991){Harjunp{\"a}{\"a}}, {Liljestrom},
  \& {Mattila}}]{1991A&A...249..493H}
{Harjunp{\"a}{\"a}}, P., {Liljestrom}, T., \& {Mattila}, K. 1991, \aap, 249,
  493

\bibitem[{{Herbig} \& {Jones}(1983)}]{1983AJ.....88.1040H}
{Herbig}, G.~H. \& {Jones}, B.~F. 1983, \aj, 88, 1040

\bibitem[{{Hilton} \& {Lahulla}(1993)}]{1993AJ....106..672H}
{Hilton}, J. \& {Lahulla}, J.~F. 1993, \aj, 106, 672

\bibitem[{{Hilton} \& {Lahulla}(1995)}]{1995A&AS..113..325H}
{Hilton}, J. \& {Lahulla}, J.~F. 1995, \aaps, 113, 325

\bibitem[{{Ho} {et~al.}(1978){Ho}, {Barrett}, \&
  {Martin}}]{1978ApJ...221L.117H}
{Ho}, P.~T.~P., {Barrett}, A.~H., \& {Martin}, R.~N. 1978, \apjl, 221, L117

\bibitem[{{H{\o}g} {et~al.}(2000){H{\o}g}, {Fabricius}, {Makarov}, {Urban},
  {Corbin}, {Wycoff}, {Bastian}, {Schwekendiek}, \&
  {Wicenec}}]{2000A&A...355L..27H}
{H{\o}g}, E., {Fabricius}, C., {Makarov}, V.~V., {et~al.} 2000, \aap, 355, L27

\bibitem[{{Huard} {et~al.}(2006){Huard}, {Myers}, {Murphy}, {Crews}, {Lada},
  {Bourke}, {Crapsi}, {Evans}, {McCarthy}, \& {Kulesa}}]{2006ApJ...640..391H}
{Huard}, T.~L., {Myers}, P.~C., {Murphy}, D.~C., {et~al.} 2006, \apj, 640, 391

\bibitem[{{Kandori} {et~al.}(2003){Kandori}, {Dobashi}, {Uehara}, {Sato}, \&
  {Yanagisawa}}]{2003AJ....126.1888K}
{Kandori}, R., {Dobashi}, K., {Uehara}, H., {Sato}, F., \& {Yanagisawa}, K.
  2003, \aj, 126, 1888

\bibitem[{{Kauffmann} {et~al.}(2005){Kauffmann}, {Bertoldi}, {Evans}, \& {the
  C2D Collaboration}}]{2005AN....326..878K}
{Kauffmann}, J., {Bertoldi}, F., {Evans}, II, N.~J., \& {the C2D
  Collaboration}. 2005, Astronomische Nachrichten, 326, 878

\bibitem[{{Kenyon} {et~al.}(1994){Kenyon}, {Dobrzycka}, \&
  {Hartmann}}]{1994AJ....108.1872K}
{Kenyon}, S.~J., {Dobrzycka}, D., \& {Hartmann}, L. 1994, \aj, 108, 1872

\bibitem[{{Kislyakov} \& {Gordon}(1983)}]{1983ApJ...265..766K}
{Kislyakov}, A.~G. \& {Gordon}, M.~A. 1983, \apj, 265, 766

\bibitem[{{Knude}(2010)}]{2010arXiv1006.3676K}
{Knude}, J. 2010, ArXiv e-prints

\bibitem[{{Knude} \& {H{\o}g}(1998)}]{1998A&A...338..897K}
{Knude}, J. \& {H{\o}g}, E. 1998, \aap, 338, 897

\bibitem[{{Knude} \& {Kaltcheva}(2010)}]{2010arXiv1003.2550K}
{Knude}, J. \& {Kaltcheva}, N. 2010, ArXiv e-prints

\bibitem[{{Kun}(1998)}]{1998ApJS..115...59K}
{Kun}, M. 1998, \apjs, 115, 59

\bibitem[{{Kun} \& {Prusti}(1993)}]{1993A&A...272..235K}
{Kun}, M. \& {Prusti}, T. 1993, \aap, 272, 235

\bibitem[{{Lee} {et~al.}(2009){Lee}, {Bourke}, {Myers}, {Dunham}, {Evans},
  {Lee}, {Huard}, {Wu}, {Gutermuth}, {Kim}, \& {Kang}}]{2009ApJ...693.1290L}
{Lee}, C.~W., {Bourke}, T.~L., {Myers}, P.~C., {et~al.} 2009, \apj, 693, 1290

\bibitem[{{Lee} \& {Myers}(1999)}]{1999ApJS..123..233L}
{Lee}, C.~W. \& {Myers}, P.~C. 1999, \apjs, 123, 233

\bibitem[{{Lee} {et~al.}(2004){Lee}, {Myers}, \& {Plume}}]{2004ApJS..153..523L}
{Lee}, C.~W., {Myers}, P.~C., \& {Plume}, R. 2004, \apjs, 153, 523

\bibitem[{{Lee} {et~al.}(1999){Lee}, {Myers}, \&
  {Tafalla}}]{1999ApJ...526..788L}
{Lee}, C.~W., {Myers}, P.~C., \& {Tafalla}, M. 1999, \apj, 526, 788

\bibitem[{{Loinard} {et~al.}(2008){Loinard}, {Torres}, {Mioduszewski}, \&
  {Rodr{\'{\i}}guez}}]{2008ApJ...675L..29L}
{Loinard}, L., {Torres}, R.~M., {Mioduszewski}, A.~J., \& {Rodr{\'{\i}}guez},
  L.~F. 2008, \apjl, 675, L29

\bibitem[{{Loinard} {et~al.}(2007){Loinard}, {Torres}, {Mioduszewski},
  {Rodr{\'{\i}}guez}, {Gonz{\'a}lez-L{\'o}pezlira}, {Lachaume}, {V{\'a}zquez},
  \& {Gonz{\'a}lez}}]{2007ApJ...671..546L}
{Loinard}, L., {Torres}, R.~M., {Mioduszewski}, A.~J., {et~al.} 2007, \apj,
  671, 546

\bibitem[{{Lombardi} {et~al.}(2010){Lombardi}, {Lada}, \&
  {Alves}}]{2010A&A...512A..67L}
{Lombardi}, M., {Lada}, C.~J., \& {Alves}, J. 2010, \aap, 512, A67+

\bibitem[{{Lynds}(1962)}]{1962ApJS....7....1L}
{Lynds}, B.~T. 1962, \apjs, 7, 1

\bibitem[{{Maheswar} {et~al.}(2010){Maheswar}, {Lee}, {Bhatt}, {Mallik}, \&
  {Dib}}]{2010A&A...509A..44M}
{Maheswar}, G., {Lee}, C.~W., {Bhatt}, H.~C., {Mallik}, S.~V., \& {Dib}, S.
  2010, \aap, 509, A44+

\bibitem[{{Maheswar} {et~al.}(2004){Maheswar}, {Manoj}, \&
  {Bhatt}}]{2004MNRAS.355.1272M}
{Maheswar}, G., {Manoj}, P., \& {Bhatt}, H.~C. 2004, \mnras, 355, 1272

\bibitem[{{Mattila}(1979)}]{1979A&A....78..253M}
{Mattila}, K. 1979, \aap, 78, 253

\bibitem[{{McCuskey}(1939)}]{1939ApJ....89..568M}
{McCuskey}, S.~W. 1939, \apj, 89, 568

\bibitem[{{Meyer} {et~al.}(1997){Meyer}, {Calvet}, \&
  {Hillenbrand}}]{1997AJ....114..288M}
{Meyer}, M.~R., {Calvet}, N., \& {Hillenbrand}, L.~A. 1997, \aj, 114, 288

\bibitem[{{Monet} {et~al.}(2003){Monet}, {Levine}, {Canzian}, {Ables}, {Bird},
  {Dahn}, {Guetter}, {Harris}, {Henden}, {Leggett}, {Levison}, {Luginbuhl},
  {Martini}, {Monet}, {Munn}, {Pier}, {Rhodes}, {Riepe}, {Sell}, {Stone},
  {Vrba}, {Walker}, {Westerhout}, {Brucato}, {Reid}, {Schoening}, {Hartley},
  {Read}, \& {Tritton}}]{2003AJ....125..984M}
{Monet}, D.~G., {Levine}, S.~E., {Canzian}, B., {et~al.} 2003, \aj, 125, 984

\bibitem[{{Morita} {et~al.}(2006){Morita}, {Watanabe}, {Sugitani}, {Itoh},
  {Uehara}, {Nagashima}, {Ebizuka}, {Hasegawa}, {Kinugasa}, \&
  {Tamura}}]{2006PASJ...58L..41M}
{Morita}, A., {Watanabe}, M., {Sugitani}, K., {et~al.} 2006, \pasj, 58, L41

\bibitem[{{Nielsen} {et~al.}(2000){Nielsen}, {J{\o}nch-S{\o}rensen}, \&
  {Knude}}]{2000A&A...358.1077N}
{Nielsen}, A.~S., {J{\o}nch-S{\o}rensen}, H., \& {Knude}, J. 2000, \aap, 358,
  1077

\bibitem[{{Onishi} {et~al.}(2002){Onishi}, {Mizuno}, {Kawamura}, {Tachihara},
  \& {Fukui}}]{2002ApJ...575..950O}
{Onishi}, T., {Mizuno}, A., {Kawamura}, A., {Tachihara}, K., \& {Fukui}, Y.
  2002, \apj, 575, 950

\bibitem[{{Perryman} \& {ESA}(1997)}]{1997ESASP1200.....P}
{Perryman}, M.~A.~C. \& {ESA}, eds. 1997, ESA Special Publication, Vol. 1200,
  {The HIPPARCOS and TYCHO catalogues. Astrometric and photometric star
  catalogues derived from the ESA HIPPARCOS Space Astrometry Mission}

\bibitem[{{Peterson} \& {Clemens}(1998)}]{1998AJ....116..881P}
{Peterson}, D.~E. \& {Clemens}, D.~P. 1998, \aj, 116, 881

\bibitem[{{Racca} {et~al.}(2009){Racca}, {Vilas-Boas}, \& {de la
  Reza}}]{2009ApJ...703.1444R}
{Racca}, G.~A., {Vilas-Boas}, J.~W.~S., \& {de la Reza}, R. 2009, \apj, 703,
  1444

\bibitem[{{Racine}(1968)}]{1968AJ.....73..233R}
{Racine}, R. 1968, \aj, 73, 233

\bibitem[{{Reipurth} \& {Gee}(1986)}]{1986A&A...166..148R}
{Reipurth}, B. \& {Gee}, G. 1986, \aap, 166, 148

\bibitem[{{Rieke} \& {Lebofsky}(1985)}]{1985ApJ...288..618R}
{Rieke}, G.~H. \& {Lebofsky}, M.~J. 1985, \apj, 288, 618

\bibitem[{{Shu} {et~al.}(1987){Shu}, {Adams}, \&
  {Lizano}}]{1987ARA&A..25...23S}
{Shu}, F.~H., {Adams}, F.~C., \& {Lizano}, S. 1987, \araa, 25, 23

\bibitem[{{Snell}(1981)}]{1981ApJS...45..121S}
{Snell}, R.~L. 1981, \apjs, 45, 121

\bibitem[{{Strai{\v z}ys} {et~al.}(1992){Strai{\v z}ys}, {Cernis},
  {Kazlauskas}, \& {Meistas}}]{1992BaltA...1..149S}
{Strai{\v z}ys}, V., {Cernis}, K., {Kazlauskas}, A., \& {Meistas}, E. 1992,
  Baltic Astronomy, 1, 149

\bibitem[{{Strai{\v z}ys} {et~al.}(2003){Strai{\v z}ys}, {{\v C}ernis}, \&
  {Barta{\v s}i{\= u}t{\.e}}}]{2003A&A...405..585S}
{Strai{\v z}ys}, V., {{\v C}ernis}, K., \& {Barta{\v s}i{\= u}t{\.e}}, S. 2003,
  \aap, 405, 585

\bibitem[{{Straizys} {et~al.}(1982){Straizys}, {Wisniewski}, \&
  {Lebofsky}}]{1982Ap&SS..85..271S}
{Straizys}, V., {Wisniewski}, W.~Z., \& {Lebofsky}, M.~J. 1982, \apss, 85, 271

\bibitem[{{Terebey} {et~al.}(2009){Terebey}, {Fich}, {Noriega-Crespo},
  {Padgett}, {Fukagawa}, {Audard}, {Brooke}, {Carey}, {Evans}, {Guedel},
  {Hines}, {Huard}, {Knapp}, {McCabe}, {Menard}, {Monin}, \&
  {Rebull}}]{2009ApJ...696.1918T}
{Terebey}, S., {Fich}, M., {Noriega-Crespo}, A., {et~al.} 2009, \apj, 696, 1918

\bibitem[{{The} {et~al.}(1994){The}, {de Winter}, \&
  {Perez}}]{1994A&AS..104..315T}
{The}, P.~S., {de Winter}, D., \& {Perez}, M.~R. 1994, \aaps, 104, 315

\bibitem[{{Tomita} {et~al.}(1979){Tomita}, {Saito}, \&
  {Ohtani}}]{1979PASJ...31..407T}
{Tomita}, Y., {Saito}, T., \& {Ohtani}, H. 1979, \pasj, 31, 407

\bibitem[{{Torres} {et~al.}(2007){Torres}, {Loinard}, {Mioduszewski}, \&
  {Rodr{\'{\i}}guez}}]{2007ApJ...671.1813T}
{Torres}, R.~M., {Loinard}, L., {Mioduszewski}, A.~J., \& {Rodr{\'{\i}}guez},
  L.~F. 2007, \apj, 671, 1813

\bibitem[{{Torres} {et~al.}(2009){Torres}, {Loinard}, {Mioduszewski}, \&
  {Rodr{\'{\i}}guez}}]{2009ApJ...698..242T}
{Torres}, R.~M., {Loinard}, L., {Mioduszewski}, A.~J., \& {Rodr{\'{\i}}guez},
  L.~F. 2009, \apj, 698, 242

\bibitem[{{van den Ancker} {et~al.}(1998){van den Ancker}, {de Winter}, \&
  {Tjin A Djie}}]{1998A&A...330..145V}
{van den Ancker}, M.~E., {de Winter}, D., \& {Tjin A Djie}, H.~R.~E. 1998,
  \aap, 330, 145

\bibitem[{{Vorobyov} \& {Basu}(2006)}]{2006ApJ...650..956V}
{Vorobyov}, E.~I. \& {Basu}, S. 2006, \apj, 650, 956

\bibitem[{{Yonekura} {et~al.}(1997){Yonekura}, {Dobashi}, {Mizuno}, {Ogawa}, \&
  {Fukui}}]{1997ApJS..110...21Y}
{Yonekura}, Y., {Dobashi}, K., {Mizuno}, A., {Ogawa}, H., \& {Fukui}, Y. 1997,
  \apjs, 110, 21

\bibitem[{{Young} \& {Evans}(2005)}]{2005ApJ...627..293Y}
{Young}, C.~H. \& {Evans}, II, N.~J. 2005, \apj, 627, 293

\bibitem[{{Young} {et~al.}(2004){Young}, {J{\o}rgensen}, {Shirley},
  {Kauffmann}, {Huard}, {Lai}, {Lee}, {Crapsi}, {Bourke}, {Dullemond},
  {Brooke}, {Porras}, {Spiesman}, {Allen}, {Blake}, {Evans}, {Harvey},
  {Koerner}, {Mundy}, {Myers}, {Padgett}, {Sargent}, {Stapelfeldt}, {van
  Dishoeck}, {Bertoldi}, {Chapman}, {Cieza}, {DeVries}, {Ridge}, \&
  {Wahhaj}}]{2004ApJS..154..396Y}
{Young}, C.~H., {J{\o}rgensen}, J.~K., {Shirley}, Y.~L., {et~al.} 2004, \apjs,
  154, 396

\bibitem[{{Zacharias} {et~al.}(2004){Zacharias}, {Urban}, {Zacharias},
  {Wycoff}, {Hall}, {Monet}, \& {Rafferty}}]{2004AJ....127.3043Z}
{Zacharias}, N., {Urban}, S.~E., {Zacharias}, M.~I., {et~al.} 2004, \aj, 127,
  3043

\end{thebibliography}

\end{document}